%% file: main.tex
\documentclass[11pt, letterpaper]{article}
\usepackage[utf8]{inputenc}
\usepackage[margin=1in]{geometry}
\usepackage{titlesec}

\titleformat{\section}
  {\large\bfseries}{\thesection}{1em}{}
\titleformat{\subsection}
  {\normalsize\bfseries}{\thesubsection}{1em}{}
\titleformat{\subsubsection}
  {\normalsize\itshape}{\thesubsubsection}{1em}{}

\usepackage{array}
\usepackage{tabularx}
\usepackage{enumitem}
\usepackage{amssymb}
\usepackage{graphicx}
\usepackage{subfigure}
\newlist{selectlist}{itemize}{2}
\setlist[selectlist]{label=$\square$,leftmargin=*,noitemsep,topsep=0pt}

\usepackage{hyperref}
\hypersetup{
    colorlinks=true,
    linkcolor=blue,
    filecolor=magenta,
    urlcolor=cyan,
}

\usepackage{acronym}

\usepackage{tikz}
\usetikzlibrary{
    arrows.meta,
    positioning,
    fit,
    calc,
    shapes.geometric,
    shapes.misc
}

\usepackage[most]{tcolorbox}

\newtcblisting{terminal}{
    enhanced,
    colback=black!3,
    colframe=black!40,
    boxrule=0.5pt,
    arc=2mm,
    left=2mm,
    right=2mm,
    top=1mm,
    bottom=1mm,
    listing only,
    listing options={
        basicstyle=\ttfamily\small,
        breaklines=true
    }
}

\titleformat{\section}[block]{\hspace{1em}\bfseries}{\thesection.}{0.5em}{}
\titleformat{\subsection}[block]{\hspace{1em}}{\thesubsection}{0.5em}{}

\input{definitions}

\newcommand{\ehandpan}[0]{DIY e-HandPan}

\begin{document}
% Create the title block
\begin{flushleft}

% Remove all text in italics when filling out the template and replace with your manuscripts corresponding text in regular font.
%\textit{Text in italics are template instructions. Remove and replace all instructions with regular font text.}

\setlength{\parindent}{0pt}
\setlength{\parskip}{10pt}
% \textbf{\large HardwareX article template}

%Insert title
%Max. 20 words. A good title should contain the fewest possible words that adequately describe the content of a paper.
\textbf{Title:} \textit{DIY e-HandPan: A new DIY Low-Cost Handpan Interface based on Arduino and ESP32 Microcontrollers}

%Insert Authors
\textbf{Authors:} \textit{Beno\^it Collin, Dominique Fourer, Eric Genotelle}

%Insert Affiliations
\textbf{Affiliations:} \textit{IBISC, University of \'Evry Paris-Saclay}

%Insert Contact Email
%Include institutional email address of the corresponding author
\textbf{Contact email:} \textit{dominique.fourer@univ-evry.fr}

%Insert Abstract
%Max. 200 words. Remember that the abstract is what readers see first in electronic abstracting and indexing services. This is the advertisement of your article. Make it interesting, and easy to be understood. Be accurate and specific, keep it as brief as possible.

\textbf{Abstract:} \textit{We present DIY e-HandPan, a new open-source, low-cost and customizable handpan audio and MIDI protocol interface designed for musical performance, education and research. 
The proposed hardware is built from inexpensive electronic components and recycled materials using widely available fabrication techniques, making it accessible to makers, educators and researchers.
The instrument can be implemented on two distinct MicroController Unit (MCU): Arduino or ESP32.
The microcontroller captures strike velocity to provide expressive musical performance comparable to that of an acoustic handpan. In addition to real-time audio and MIDI generation, DIY e-HandPan integrates a bi-color LED guidance system capable of displaying musical sequences from MIDI files, providing an effective learning aid for beginners and educational activities. 
The modular architecture allows users to easily customize the number of notes, hardware configuration and embedded software according to specific applications. 
We present the complete hardware design, firmware and assembly instructions and we discuss the design choices and limitations, to evaluate the system in representative educational and musical performance scenarios.
All design files, source code and documentation are released under an open-source license to improve the reproducibility and encourage further developments by the open hardware community.
}

%Insert Keywords
% At least 3 keywords. There is no limit on the no. of keywords you can list. Please remember that effective keywords should not repeat words appearing in your title, and should be neither too general nor too narrow.
%\textbf{Keywords:} \textit{HandPan, music performance, MIDI, Low-cost, DIY, Microcontroller, Arduino, ESP32}

\textbf{Specifications table:}

%\tabulinesep=1ex
\begin{tabularx}{\linewidth}{|X|X|}
\hline  \textbf{Hardware name} & \textit{DIY e-HandPan}
  %Please specify the name of the hardware that you invented / customized
  \\
  \hline \textbf{Subject area} & %
  % Please state the subject area most relevant to the original community for which this hardware was developed. Example subject areas are listed below.
  \begin{itemize}
  %\item \textit{Engineering and Material Science}
  %\item \textit{Chemistry and Biochemistry}
  %\item \textit{Medical (e.g. Pharmaceutical Science)}
  %\item \textit{Neuroscience}
  %\item \textit{Biological Sciences (e.g. Microbiology and Biochemistry)}
  %\item \textit{Environmental, Planetary and Agricultural Sciences}
  \item \textit{Educational Tools and Open Source Alternatives to Existing Infrastructure}
  %\item \textit{General}
  \end{itemize}
  \\
  \hline \textbf{Hardware type} &
  \begin{itemize}
  \item \textit{Electronic musical instrument}
  %\item \textit{Imaging tools}
  %\item \textit{Measuring physical properties and in-lab sensors}
  %\item \textit{Biological sample handling and preparation}
  %\item \textit{Field measurements and sensors}
  \item \textit{Electrical engineering and computer science}
  %\item \textit{Mechanical engineering and materials science}
  %\item \textit{Other (please specify)}
  \end{itemize}
  \\
\hline \textbf{Open source license} &
  %Please specify the open source license. For more details see the guide to authors.
  %Please specify the open source license. For more details see the guide to authors.
  %\textit{Creative Common Licence}
\begin{itemize}
\item \textit{MIT License for the Arduino and ESP32 firmware}
\item \textit{CERN Open Hardware Licence Version 2 -- Permissive (CERN-OHL-P-2.0) for the hardware design files}
\end{itemize}
  \\
\hline \textbf{Cost of hardware} &
  %Approximate cost of hardware (complete breakdown will be included in the Bill of Materials).
  %\textit{Approximate cost of hardware (complete breakdown will be included in the Bill of Materials).}
  \textit{Approximately 80 USD}
  \\
\hline \textbf{Source file repository} &
  % Link to the source file repository
  \url{https://github.com/dfourer/eHandPan}
  %\textit{DOI URL to an approved source file repository: \href{https://data.mendeley.com/}{Mendeley Data}, the \href{http://osf.io}{OSF}, or \href{https://zenodo.org/}{Zenodo} \href{https://doi.org/10.5281/zenodo.3346799}{(instructions)}. For example:} \url{https://doi.org/10.5281/zenodo.3346799}
\\\hline
\end{tabularx}

\end{flushleft}
% create the main body of the paper
\clearpage

\section{Hardware in context}
% Include a short description of the hardware, putting into context of similar open hardware and proprietary equipment in the field.
%\textit{Include a short description of the hardware, putting into context of similar open hardware and proprietary equipment in the field.}

\input{introduction}

\section{Hardware description} \label{sec:hardware}

\input{hardware}
\section{Software description} \label{sec:software}

\input{software}

%% TODO : Benoit
\section{Build instructions}  \label{sec:building}

\input{implementation}

\section{Operation instructions} \label{sec:operating}

Before operating the instrument, verify that all internal wires are insulated, that the DB25 connector is correctly secured when applicable, and that no conductive part can move freely inside the salad bowl.

\subsection{Arduino version}

The Arduino version is operated as follows:

\begin{enumerate}
    \item connect the Arduino board to the host computer through USB;
    \item verify that the board is detected as a USB-MIDI device;
    \item open the selected digital audio workstation;
    \item create or select a MIDI track receiving data from the MocoLUFA device;
    \item load a compatible virtual instrument or sampler;
    \item strike each compact disc and verify that the expected MIDI note is received;
    \item when LED guidance is used, route the MIDI output of the digital audio workstation back to the MocoLUFA device;
    \item verify that the LED corresponding to each transmitted MIDI note is illuminated.
\end{enumerate}

The instrument should be struck using the fingers or hands only. Hard sticks or metallic objects may damage the compact discs, piezoelectric sensors or adhesive joints.

A MIDI feedback loop must be avoided. The MIDI output sent from the host to the LEDs must not be routed back repeatedly into the same MIDI track. Such a loop may generate repeated notes, continuous LED activation or excessive MIDI traffic.

\subsection{ESP32 version}

The ESP32 version is operated as follows:

\begin{enumerate}
    \item verify that the DB25 cable is fully inserted and secured;
    \item connect the audio output and any required external loudspeaker or amplifier;
    \item power the ESP32 interface using the specified supply;
    \item wait for completion of the startup procedure;
    \item connect to the network interface when remote configuration or learning mode is required;
    \item select the desired operating mode;
    \item strike each note area and verify correct note detection, LED response and audio playback.
\end{enumerate}

The DB25 connector must not be inserted or removed while the system is powered. The instrument must be disconnected immediately if abnormal heating, intermittent operation, unexpected LED behavior or electrical noise is observed.

\section{Validation and characterization} \label{sec:validation}

\input{validation}

\section{Capabilities and limitation}

The proposed hardware offers the following capabilities:

\begin{itemize}
\item Standalone electronic handpan (ESP32 implementation).
\item USB-MIDI controller compatible with standard DAWs.
\item Velocity-sensitive note detection.
\item Polyphonic sample playback.
\item Interactive learning mode based on MIDI files.
\item Browser-based wireless configuration.
\item Open-source firmware for Arduino and ESP32.
\item Low-cost construction using commercially available components.
\end{itemize}

The current implementation has the following limitations:

\begin{itemize}
\item Mechanical construction quality influences the sensitivity and playing consistency of the instrument.
\item Impact thresholds must be adjusted according to the selected piezoelectric sensors.
\item The Arduino implementation requires an external synthesizer.
\item The ESP32 implementation relies on prerecorded samples rather than physical sound synthesis.
\end{itemize}

Overall, the experimental evaluation demonstrates that the proposed hardware provides reliable impact detection, adequate electrical protection of the microcontroller inputs, and stable operation for both the Arduino and ESP32
implementations. Together with the educational deployments, these experiments confirm that the \ehandpan{} is a reproducible and low-cost platform for digital musical instrument research, teaching, and performance.

\section{Acknowledgements}

The authors would like to thank Raymond Herren and Jean-Robert Basoin from CEA Paris-Saclay for their significant contribution to the design and validation of the piezoelectric signal-conditioning circuit.

The authors also gratefully acknowledge Professor Pierre Couprie from the University of Evry Paris-Saclay (CHCSC, EA 2448) for his valuable support and for integrating the \ehandpan{} into educational activities and teaching projects.

\section{Declaration of interest}

The authors declare that they have no known competing financial interests or personal relationships that could have influenced the work reported in this paper.

\bibliographystyle{elsarticle-num}
\bibliography{biblio}

\end{document}

%% file: definitions.tex
\acrodef{diy}[DIY]{Do It Yourself}
\acrodef{mcu}[MCU]{Microcontroller}
\acrodef{dmi}[DMI]{Digital Music Instrument}
\acrodef{midi}[MIDI]{Music Instrument Digital Interface}
\acrodef{vr}[VR]{Virtual Reality}
\acrodef{pcb}[PCB]{Printed Circuit Board}

\newcommand{\etal}[0]{\textit{et al.} }

%% file: introduction.tex
The rise of digital music technologies has opened new possibilities for designing cost-effective, customizable and versatile musical interfaces. Recent advances in embedded electronics and open-source microcontroller platforms have enabled the development of \acp{dmi} that are affordable, reproducible and easily adaptable to a wide range of musical applications.

Among contemporary percussion instruments, the handpan, also known as the \emph{Hang drums}, stands out for its rich harmonic spectrum, intuitive playing technique and remarkable expressive capabilities. It is an idiophone instrument introduced in the early 2000s \cite{wong2023history}, whose popularity has steadily increased owing to its distinctive meditative timbre and its suitability for both solo performance and music therapy applications \cite{alon2015analysis}. Traditionally, a handpan consists of two deep-drawn steel shells glued together to form a resonant cavity. Depending on its design, the instrument may be diatonic or chromatic and typically provides between 6 and 14 playable notes distributed over the shell. %shell

Despite its growing popularity, access to the instrument remains limited because high-quality handpans are handcrafted and therefore relatively expensive, with commercial instruments typically costing several hundred to several thousand euros. Such prices are a significant barrier for beginners, educational institutions and makers interested in experimenting with the instrument. Furthermore, because the handpan is a relatively recent instrument, qualified teachers and pedagogical resources remain comparatively limited.

Several systems try to reproduce or to augment the handpan playing experience. Gosling \etal proposed a \ac{vr} environment that simulates both the instrument and its expressive behaviour \cite{gosling23}. Although immersive, this approach requires dedicated \ac{vr} hardware and a powerful computer, limiting its accessibility. Commercial hardware solutions of electronic handpan are also available, while DIY projects such as Panduino demonstrate the feasibility of \ac{midi} handpan controllers. However, as summarized in Table~\ref{tab:handpan_hardware}, existing solutions either remain proprietary and relatively expensive or are not publicly documented, preventing their reproduction and further development by the community.

More generally, the emergence of open-source embedded platforms such as Arduino \cite{arduino2015arduino,monk2016programming} and ESP32 \cite{babiuch2019using,cameron2023esp32} has considerably simplified the development of expressive \acp{dmi}. Previous studies have demonstrated that low-cost electronics can successfully capture expressive gestures through velocity-sensitive sensing, haptic interaction and embedded processing \cite{Overholt2009,Berdahl2010}. These advances have contributed to democratizing \ac{dmi} development while promoting reproducible and customizable hardware solutions. Nevertheless, most existing systems are generic controllers and do not specifically reproduce the ergonomic layout and playing technique of a handpan.

To address these limitations, we present \textbf{DIY e-HandPan}, an open-source, low-cost and fully customizable handpan \ac{midi} interface designed using readily available electronic components and recycled materials. The proposed instrument preserves one of the essential characteristics of acoustic handpans by measuring the striking velocity, thereby enabling expressive musical performance. It communicates through the standard \ac{midi} protocol and can therefore be connected to any software or hardware synthesizer. In addition, an integrated bi-color LED guidance system allows beginners to visualize notes while learning musical pieces directly from \ac{midi} files.

\clearpage

Compared with existing solutions, \textbf{\ehandpan{}} offers several advantages:
\begin{itemize}
\item a complete hardware cost below 100~USD;
\item fully open-source hardware, firmware and assembly documentation;
\item straightforward construction using commonly available components and simple fabrication techniques;
\item customizable layouts that can be adapted to different handpan scales and educational scenarios;
\item an integrated visual learning system designed for music education and demonstrations.
\end{itemize}

To the best of our knowledge, \textbf{\ehandpan{}} is the first fully documented open-source handpan \ac{midi} controller combining velocity-sensitive interaction, educational visual feedback and a complete set of reproducible hardware design files.

The complete hardware documentation, firmware and assembly instructions are made publicly available to ensure reproducibility and to encourage further developments by the open-source hardware community.
The paper is organized as follows. In Sections~\ref{sec:hardware} and \ref{sec:software}, we respectively detail the hardware and firmware of both versions (arduino and ESP32) of the \ehandpan{}.
The building and operating instructions are respectively presented in Sections~\ref{sec:building} and \ref{sec:operating}. Finally, we discuss and validate the implementation choices in Section~\ref{sec:validation}.

% \begin{table}[t]
% \centering
% \caption{Comparison of existing handpan hardware systems with the new proposed instrument.}
% \label{tab:handpan_hardware}
% \resizebox{\textwidth}{!}{%
% \begin{tabular}{|l|l|l|c|c|c|c|r|}
% \hline
% \textbf{System} &
% \textbf{Project} &
% \textbf{Ergonomy} &
% \textbf{Chromatic full octave} &
% \textbf{Notes} &
% \textbf{Open HW} &
% \textbf{Velocity} &
% \textbf{Cost (USD)} \\
% \hline

% Acoustic Handpan &
% Commercial &
% 600 x 140 + 48 x 110mm &
% Yes &
% 14 &
% No &
% Yes &
% 2500 for 2 items \\
% \hline

% Lumen Handpan \cite{lumen} &
% Commercial &
% 600 x 200mm  &
% with 2 items if configurable &
% 9 &
% No &
% Yes &
% 1,729 \\
% \hline

% Neotone \cite{neotone,neotone_demo} &
% Commercial &
% 470 x 160mm &
% with 2 items if configurable &
% 10 &
% No &
% Yes &
% 2,330--3,690 \\
% \hline

% Roland Mood Pan \cite{roland_moodpan,roland_moodpan_demo} &
% Commercial &
% 316 x 94mm &
% with 2 items if configurable &
% 9 &
% No &
% Yes &
% 730 \\
% \hline

% Panduino \cite{panduino13,panduino16,panduino17} &
% DIY (closed) &
% about 230 x 50mm  &
% if configurable &
% 13/16/17 &
% No &
% Yes &
% N/A \\
% \hline

% \textbf{DIY eHandPan (proposed)} &
% \textbf{Open-source} &
% \textbf{450 x 90mm + 40 x 70mm} &
% \textbf{Yes} &
% \textbf{8 + 6} &
% \textbf{Yes} &
% \textbf{Yes} &
% \textbf{$<100$} \\
% \hline
% %
% \end{tabular}}
% \end{table}

\begin{table}[t]
\centering
\caption{Comparison of existing handpan hardware systems.}
\label{tab:handpan_hardware}
\resizebox{\textwidth}{!}{\begin{tabular}{|l|c|c|c|c|c|c|r|}
\hline
\textbf{System} &
\textbf{Type} &
\textbf{Dimensions (mm)} &
\textbf{Chromatic} &
\textbf{Notes} &
\textbf{Open} &
\textbf{Velocity} &
\textbf{Cost (USD)}\\
\hline
Acoustic Handpan &
Comm. &
560$\times$130 + 48$\times$110 &
2 units &
8+6 &
No &
Yes &
2500\\
\hline
Lumen &
Comm. &
600$\times$200 &
Optional &
9 &
No &
Yes &
1729\\
\hline
Neotone &
Comm. &
470$\times$160 &
Optional &
10 &
No &
Yes &
2330--3690\\
\hline
Roland Mood Pan &
Comm. &
316$\times$94 &
Optional &
9 &
No &
Yes &
730\\
\hline
Panduino &
DIY &
230$\times$50 &
Config. &
13/16/17 &
No &
Yes &
N/A\\
\hline
\textbf{\ehandpan{}} &
\textbf{Open} &
\textbf{450$\times$130 + 40$\times$100} &
\textbf{Yes} &
\textbf{8+6} &
\textbf{Yes} &
\textbf{Yes} &
\textbf{$<100$}\\
\hline
\end{tabular}}
\end{table}

%% file: hardware.tex
\subsection{Mechanical Design}

\begin{figure}[!ht]
\centering
\subfigure[Arduino Mega version]{\includegraphics[width=0.45\textwidth]{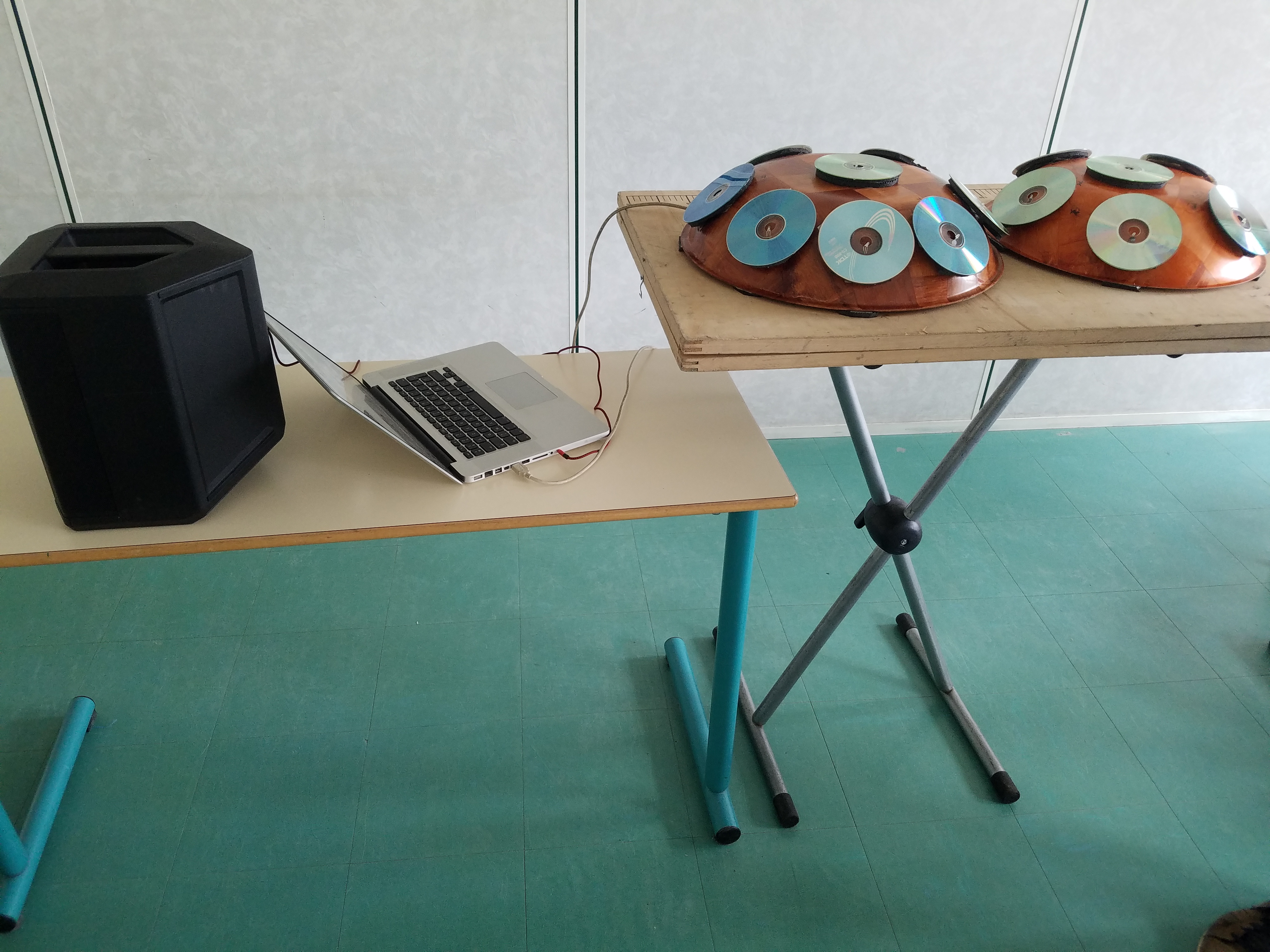}}
\hfill
\subfigure[ESP32-S3 autonomous version]{\includegraphics[width=0.45\textwidth]{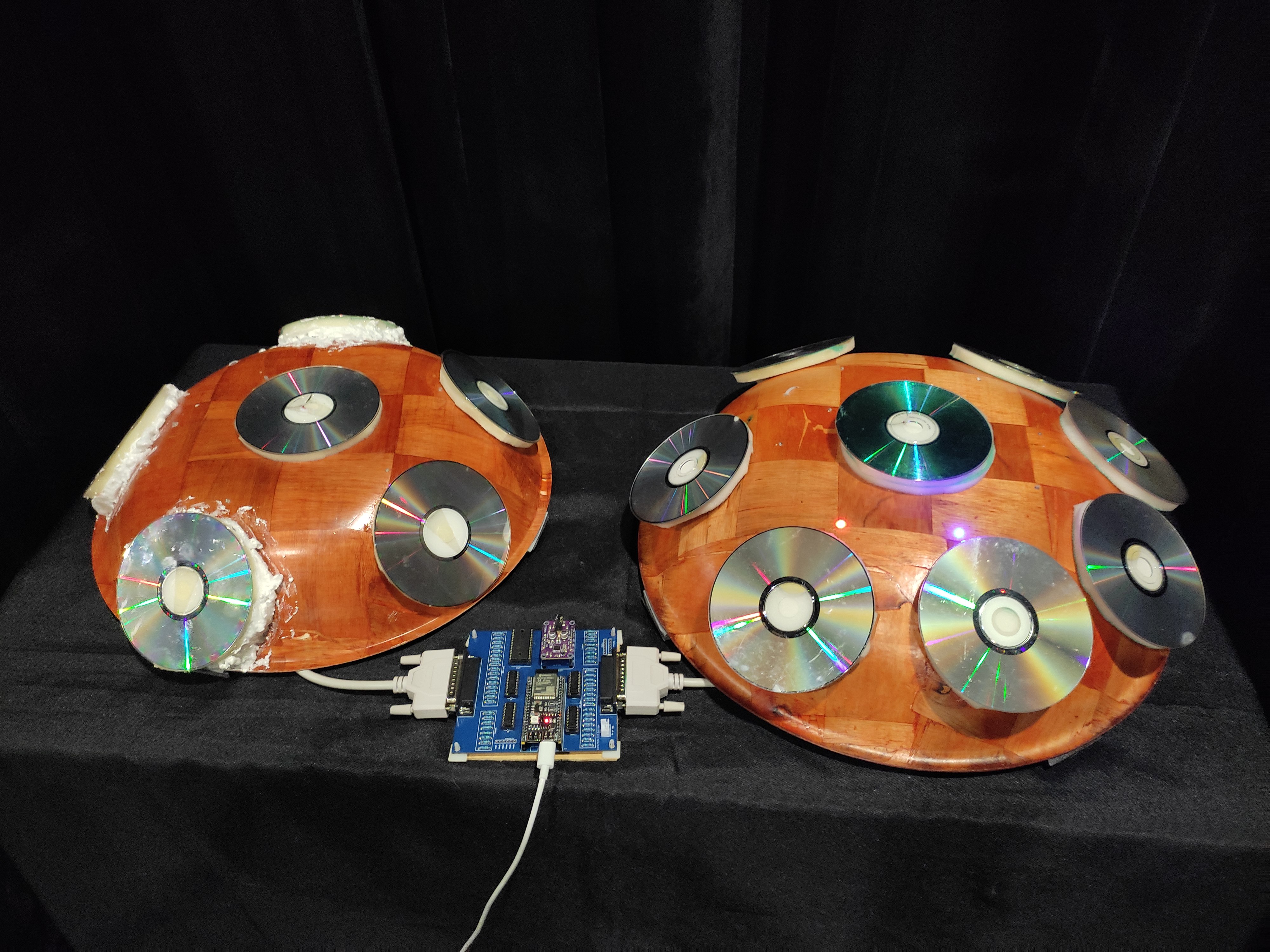}}
\caption{Two versions of the DIY e-HandPan. Left: the original MIDI controller based on an Arduino Mega. Right: the autonomous version based on an ESP32-S3 with embedded audio playback and wireless connectivity.}
\label{fig:handpan}
\end{figure}

The DIY eHandPan is designed as a modular open-hardware instrument that can be assembled in several configurations ranging from 6 to 14 notes (cf. Fig.~\ref{fig:handpan}) similarly to the original acoustic instrument.
In the present work, we propose 2 different complementary electronic implementations:
\begin{itemize}
    \item The first implementation relies on an Arduino Uno or Arduino Mega \cite{banzi2009getting} and behaves as a standard MIDI controller. It targets educational activities, rapid prototyping and experimental developments where the instrument is connected to an external synthesizer or a computer through the MIDI protocol. Owing to the large number of analog inputs available on Arduino boards, this version can be assembled without a dedicated printed circuit board or additional interface electronics, making it particularly suitable for makers who aim to customize the hardware.
    \item The second implementation is based on an ESP32-S3 microcontroller \cite{cameron2023esp32} and provides a fully autonomous electronic instrument. In addition to impact detection, it embeds polyphonic playback of prerecorded handpan samples, wireless communication and an interactive learning interface. This implementation requires a dedicated printed circuit board incorporating analog multiplexers and shift registers to minimize the number of microcontroller I/O pins while supporting the complete 14-note instrument.
\end{itemize}

Table~\ref{tab:materials} summarizes the mechanical and electronic components shared by both implementations. The instrument body is built from two commercially available wooden salad bowls. Each playing surface consists of a recycled compact disc mechanically coupled to a piezoelectric sensor through a foam spacer. A bi-color LED located on each note provides visual feedback during educational modes while preserving the playing ergonomics of a conventional handpan.

Recycled compact discs were selected as playing surfaces because they provide a rigid, lightweight and inexpensive circular support while simplifying the integration of both the piezoelectric sensor and the bi-color LED. In addition to their widespread availability as discarded consumer products, compact discs offer highly reproducible dimensions and sufficient mechanical rigidity to ensure consistent impact detection. Several alternative supports, including plastic containers and yogurt lids, were evaluated during the design phase; however, recycled compact discs provided the best compromise between mechanical robustness, manufacturing simplicity, sustainability and playing comfort.
The additional components specific to each implementation are listed in Table~\ref{tab:bom_arduino} for the Arduino version and Table~\ref{tab:bom_esp32} for the ESP32 implementation.

\begin{table}[t]
\centering
\caption{Common bill of materials required to build a 14-note DIY eHandPan for both Arduino and ESP32 version. Unit prices are indicative.}
\label{tab:materials}
\begin{tabular}{|l|l|c|c|}
\hline
\textbf{Component} & \textbf{Specification} & \textbf{Qty.} & \textbf{Unit price (USD)} \\
\hline
Wooden salad bowls                      & 40--45 cm diameter        & 2     & 15 \\
Recycled compact discs                  & One per note              & 14    & Recycled \\
Foam pads                               & Sensor support            & 14    & 0.2 \\
Piezoelectric sensors                   & 27 mm                     & 14    & 1.8 \\
Bi-color LEDs                           & Red/Green, common cathode & 14    & 0.25 \\
Electronic wire                         & AWG26 (sensor wiring)     & 2 m   & 3.7 \\
%Heat-shrink tubing & 3.2 mm & 10 m & 10.1 \\
USB-C power connector                   & Panel mount               & 1     & 3.2 \\
\hline
R1: 1 M$\Omega$ resistor & Piezo bias resistor          & 14 & 0.02 \\
16 V Zener diode & Piezo overvoltage protection         & 14 & 0.08 \\
Signal diode &  1N4148 or equivalent for piezo sensor   & 14 & 0.03 \\
%100 $\Omega$ resistor & LED protection resistor         & 14 & 0.02 \\
\hline
\end{tabular}
\end{table}

\subsection{Piezoelectric sensing circuit}\label{sec:conditioning}

\begin{figure}[!ht]
\centering
\includegraphics[width=0.5\textwidth]{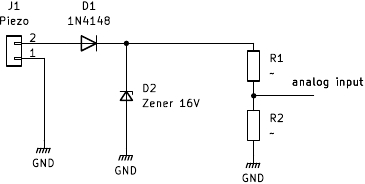}
\caption{Conditioning circuit of a single piezoelectric sensor. The piezoelectric sensor is interfaced to the microcontroller (Arduino or ESP32) analog input through a voltage conditioning stage composed of two protection diodes and two resistors. This circuit protects the microcontroller
while preserving the signal dynamics required for MIDI velocity estimation.}
%The circuit attenuates and clamps the voltage generated by the piezoelectric transducer before it is applied to the analog input, while preserving the signal dynamics required for MIDI velocity estimation.}
\label{fig:piezo_conditioning}
\end{figure}

Each note is equipped with an identical signal-conditioning circuit illustrated in Fig.~\ref{fig:piezo_conditioning}. 
The circuit performs two main functions: (i) protecting the analog input against positive and negative voltage transients, and (ii) scaling the sensor output to the operating voltage range while preserving the signal dynamics required for MIDI velocity estimation.

The protection stage combines a Zener diode and a conventional diode. The Zener diode clamps positive voltage peaks generated by strong impacts, whereas the conventional diode protects the input against negative voltage excursions. Together, these two components prevent excessive voltages from reaching the analog input and improve the long-term reliability of the acquisition electronics.

The resistor network acts as a voltage divider that attenuates the voltage generated by the piezoelectric sensor before it reaches the analog input. Assuming the input impedance of the microcontroller is much larger than the divider resistance, neglecting the conduction of the protection diodes, the input voltage is given by:
\begin{equation} 
V_{\mathrm{in}} = V_{\mathrm{piezo}} \frac{R_2}{R_1+R_2}, \label{eq:voltage_divider} 
\end{equation}
where $V_{\mathrm{piezo}}$ is the voltage produced by the piezoelectric sensor and $V_{\mathrm{in}}$ is the voltage applied to the microcontroller analog input. 
The value of $R_2$ is selected so that the maximum expected impact voltage remains within the operating range of the target microcontroller while preserving sufficient dynamic range for MIDI velocity estimation. The conditioning circuit is common to both hardware implementations. Only the value of resistor $R_2$ differs according to the analog input voltage of the target platform: a value of 470~k$\Omega$ is used for the Arduino implementation operating from a 5~V reference, whereas a value of 200~k$\Omega$ is selected for the ESP32 implementation operating at 3.3~V.

A simplified conditioning circuit using only the series resistor $R_1=1M\Omega{}$ has also been evaluated. Although this alternative reduces the component count and operates correctly under moderate playing conditions, it does not provide adequate protection against the large voltage transients that may be generated by repeated or vigorous impacts on the piezoelectric sensor. For this reason, the protected circuit shown in Fig.~\ref{fig:piezo_conditioning} is recommended for long-term operation and is used throughout the present work. The implementation choices and the chosen values of the resistors are further discussed and assessed in Section~\ref{sec:validation}.

\subsection{Arduino implementation: MIDI controller}
\begin{figure*}[!ht]
\centering
\includegraphics[width=\textwidth]{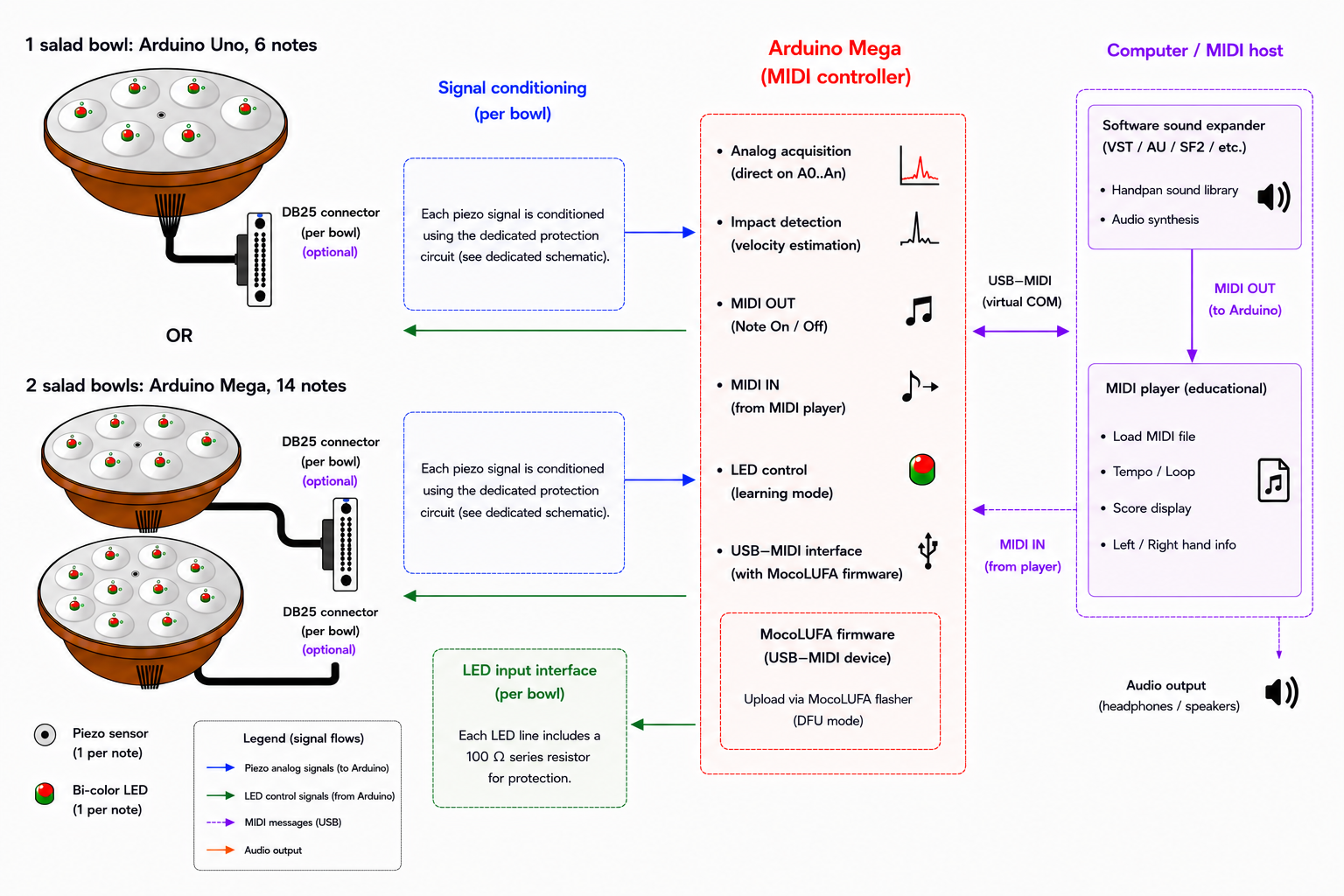}
\caption{Hardware and functional architecture of the Arduino-based DIY
eHandPan. Each wooden bowl can be connected through a DB25, RJ45 or any cable carrying up to
8 piezoelectric signals (RJ45) with a shared ground. After signal conditioning,
the sensors are directly connected to the Arduino analog inputs, without using
an analog multiplexer.}
\label{fig:arduino_architecture}
\end{figure*}
\begin{table}[!ht]
\centering
\caption{Bill of materials specific to the Arduino implementation. Unit prices are indicative.}
\label{tab:bom_arduino}
\begin{tabular}{|l|l|c|c|}
\hline
\textbf{Component} & \textbf{Specification} & \textbf{Qty.} & \textbf{Unit price (USD)} \\
\hline
Arduino Mega 2560 & Recommended microcontroller & 1 & 60 \\
Arduino Uno R3 & Alternative for reduced versions & 1 & 40 \\
USB cable & USB-A to USB-B & 1 & 3 \\
RJ11, RJ45 or DB25 cable & Sensor wiring harness (optional) & 1 & 3 \\
220 $\Omega$ resistor & LED protection resistor         & 14 & 0.02 \\ % 100 $\Omega$
R2: 470 k$\Omega$ resistor & Input voltage conditioning & 14 & 0.02 \\
\hline
\end{tabular}
\end{table}

The Arduino implementation (cf. Fig.~\ref{fig:arduino_architecture} and Table~\ref{tab:bom_arduino}) provides the simplest and most accessible version of the DIY eHandPan. It operates as a standard \ac{midi} controller and therefore requires an external synthesizer, software instrument, or digital audio workstation to generate the musical sound. Although less integrated than the ESP32 implementation, the Arduino version remains valuable because it can be assembled without a custom PCB, relies exclusively on widely available development boards, and provides direct access to every acquisition channel. It therefore constitutes an ideal platform for experimentation, education, and rapid prototyping of new sensing strategies.

Unlike the ESP32 implementation, the Arduino architecture was not designed as a standalone electronic instrument but as an open and modular hardware platform. Since the Arduino Uno and Mega provide respectively 8 and 16 analog inputs, the conditioned piezoelectric signals can be connected directly to dedicated analog inputs without requiring analog multiplexers or additional interface electronics. This simplifies both the acquisition hardware and firmware while allowing each sensor to be sampled independently with minimal latency.

As a result, the complete instrument can be assembled using only commercially available modules and point-to-point wiring, eliminating the need for a dedicated \ac{pcb}. This considerably simplifies replication while facilitating hardware modifications and extensions by makers, students, and educators.

The firmware continuously acquires the conditioned piezoelectric signals, detects impacts, estimates their velocity, and converts each strike into MIDI Note On and Note Off messages. Since the firmware is only responsible for sensor acquisition, MIDI generation, and LED control, the computational load remains low and is fully compatible with both the Arduino Uno and Arduino Mega. During development, the firmware is first validated through the standard USB serial interface. Once the acquisition has been verified, the USB interface firmware is replaced by MocoLUFA, allowing the board to appear as a native class-compliant USB-MIDI device. The instrument can then be connected directly to an external synthesizer, digital audio workstation, or any MIDI-compatible software without requiring an external serial-to-MIDI bridge.

Besides generating MIDI events, the firmware also controls the bi-color LEDs located beneath each playing surface. These LEDs provide immediate visual feedback during performance and facilitate debugging by confirming impact detection. They also support an interactive learning mode in which MIDI files are translated into illuminated note sequences displayed directly on the instrument.

The components required for this implementation are listed in Table~\ref{tab:bom_arduino}. Although the Arduino Uno is sufficient for 6- and 8-note instruments, the Arduino Mega is recommended because its 16 analog inputs allow the implementation of the complete 14-note configuration without analog multiplexing.

Compared with the ESP32 implementation, the Arduino architecture offers fewer integrated features but provides a simpler, lower-cost, and highly modular hardware platform that can be assembled without a custom PCB and easily modified using standard prototyping tools. For these reasons, it remains particularly well suited for education, experimentation, and the rapid evaluation of new hardware and firmware developments.

\subsection{ESP32 implementation: Autonomous instrument (audio + MIDI)}
\begin{figure*}[ht!]
\centering
\includegraphics[width=\textwidth]{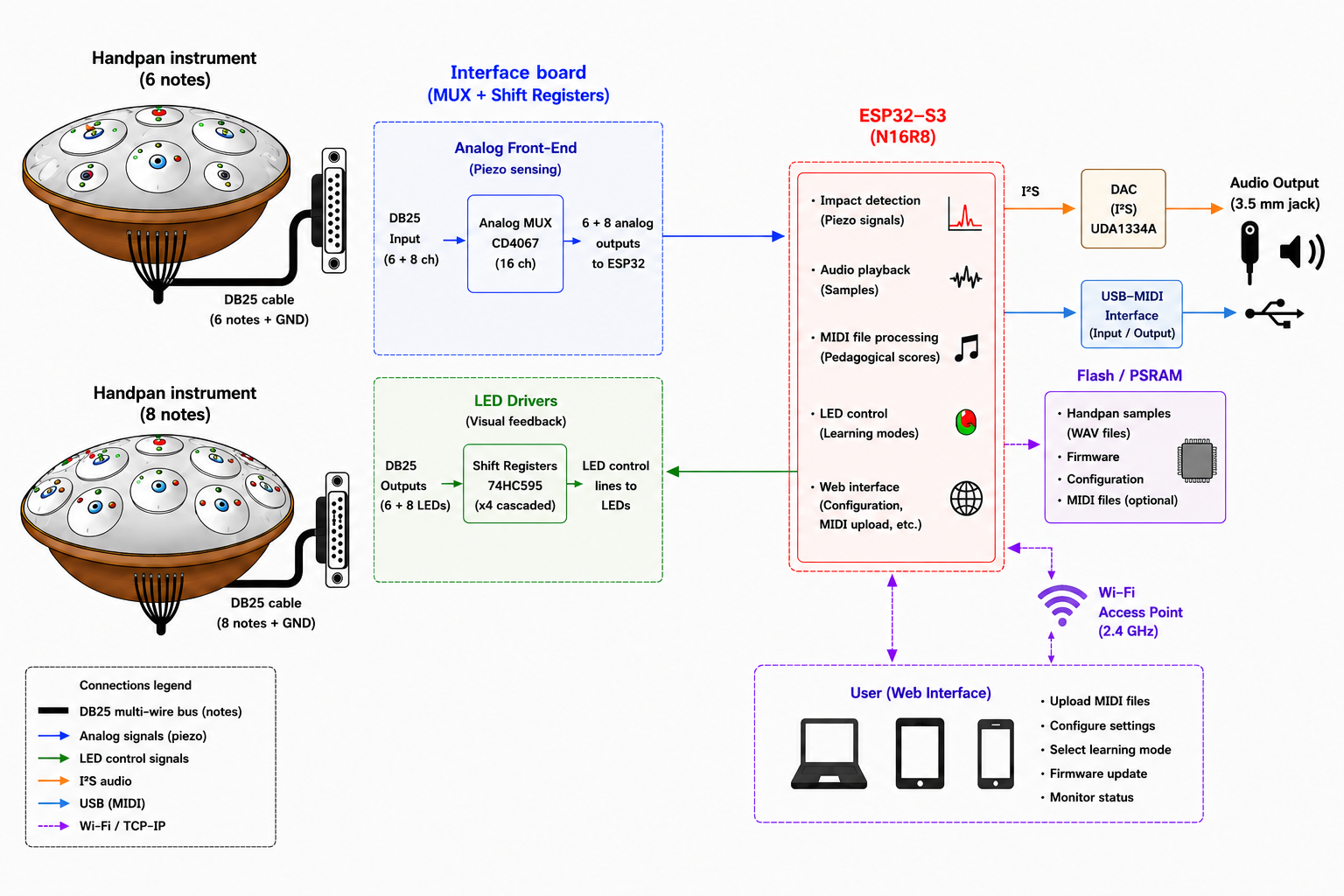}
\caption{Hardware and functional architecture of the autonomous DIY eHandPan.
The 14 piezoelectric sensors and bi-color LEDs are connected through a DB25
cable to a dedicated interface board containing an analog multiplexer and
cascaded shift registers. The ESP32-S3 performs impact detection, polyphonic playback of prerecorded handpan samples, MIDI score processing, LED control, and wireless communication
with a browser-based user interface.}
\label{fig:esp32_architecture}
\end{figure*}

\begin{table}[!ht]
\centering
\caption{Bill of materials specific to the ESP32 implementation. Unit prices are indicative.}
\label{tab:bom_esp32}
\begin{tabular}{|l|l|c|c|}
\hline
\textbf{Component} & \textbf{Specification} & \textbf{Qty.} & \textbf{Unit price (USD)}\\
\hline
ESP32-S3 development board & N16R8 & 1 & 8.2\\
I$^2$S DAC module & UDA1334A & 1 & 4.1\\
Analog multiplexer & CD4067 (16 channels) & 1 & 5.9\\
Shift register & 74HC595 & 4 & 1.8\\
16-pin DIP socket & For CD4067 & 1 & 1.8\\
24-pin DIP socket & For ESP32 module & 1 & 1.8\\
40-pin IC socket & Expansion connector & 1 & 2.3\\
DB25 connector & PCB mount & 1 & 2.9\\
JST-XH connectors & 3-pin sensor cables & 14 & 0.25\\
Potentiometer & Linear (volume) & 1 & 4.7\\
220~$\Omega$ resistors & LED current limiting & 14 & 0.02\\
200~k$\Omega$ resistors & Piezo conditioning ($R_2$) & 14 & 0.02\\
\hline
\end{tabular}
\end{table}

The ESP32 implementation (cf. Fig.~\ref{fig:esp32_architecture} and Table~\ref{tab:bom_esp32}) provides a fully autonomous version of the DIY eHandPan. 
In contrast to the Arduino-based design, it does not require an external computer or software synthesizer during performance. Instead of only transmitting MIDI messages, the instrument directly plays prerecorded handpan samples stored in its internal Flash memory and generates an analog audio signal through a standard 3.5~mm output.

This implementation is intended to operate as a complete electronic musical instrument rather than primarily as a development platform. In addition to real-time audio playback, it integrates wireless configuration, a browser-based control interface, MIDI score playback, and interactive learning functions, while retaining the same mechanical construction and playing ergonomics as the Arduino implementation.

Our proposed system is built around an ESP32-S3-N16R8 microcontroller board equipped with a dual-core 240~MHz processor, integrated Wi-Fi, 8~MB of PSRAM, and 16~MB of Flash memory. 
The ESP32-S3 was selected because it simultaneously provides sufficient processing power for real-time audio synthesis, integrated Wi-Fi for browser-based configuration, and enough memory to store multiple sampled instruments. These resources allow the firmware to perform sensor acquisition, impact detection, polyphonic sample playback, LED control, MIDI processing, and web-server operation concurrently.
However, the ESP32-S3 does not provide enough directly usable analog inputs to independently acquire the 14 piezoelectric sensors while simultaneously supporting the remaining peripherals of the instrument, including the integrated Wi-Fi interface, audio subsystem, LEDs, and user controls.
%On the implemented system, 18 ADC input are available and 10 are unavailable when Wi-Fi interface is active.

A 16-channel CD4067 analog multiplexer is therefore used to sequentially route the conditioned sensor signals to a single analog-to-digital converter input. The active multiplexer channel is selected using four digital address lines (A--D), allowing the 14 piezoelectric sensors to share a single ADC input. Since only one or a few notes are typically struck simultaneously during performance, sequential acquisition provides sufficient temporal resolution while considerably reducing the number of required ADC inputs.

Similarly, four cascaded 74HC595 shift registers are used to control the 28 LED channels (two colors for each of the 14 bi-color LEDs). Thanks to the cascaded architecture, all LED channels are controlled using only three digital signals (DATA, CLOCK, and LATCH), thereby considerably reducing the number of required GPIO pins. These additional components increase the electronic complexity compared with the Arduino implementation but enable all sensing and control functions to be integrated into a dedicated printed circuit board.

%  A 16-channel CD4067 analog multiplexer is therefore used to route the conditioned sensor signals sequentially to a single analog-to-digital converter input. Since only one or a few notes are typically struck simultaneously during performance, sequential acquisition provides sufficient temporal resolution while considerably reducing the number of required ADC inputs. Similarly, 4 cascaded 74HC595 shift registers independently control the 14 bi-color LEDs while significantly reducing the number of GPIO pins required. These additional components increase the electronic complexity compared with the Arduino implementation but allow all sensing and control functions to be integrated on a dedicated printed circuit board.

Audio synthesis is performed by triggering prerecorded handpan samples stored in Flash memory. Multiple samples can be reproduced simultaneously, allowing natural note overlap during performance. The resulting digital audio stream is transmitted through the ESP32 I$^2$S peripheral to an external UDA1334A digital-to-analog converter. The DAC provides low-latency polyphonic audio output for headphones or an amplified loudspeaker connected to the 3.5~mm audio jack. A potentiometer mounted on the interface board provides direct volume adjustment. Although primarily intended as a standalone instrument, the ESP32 implementation can also operate as a class-compliant USB-MIDI controller compatible with external synthesizers and digital audio workstations.

The embedded firmware also configures the ESP32 as a Wi-Fi access point and hosts an HTTP server. No Internet connection is required, since users connect directly to the wireless network created by the instrument. The eHandPan can therefore be configured from a smartphone, tablet, or computer using any standard web browser, without requiring a dedicated application. The web interface allows users to select instrument parameters, adjust playback settings, upload MIDI files, and activate the educational functions.

As in the Arduino implementation, a red/green bi-color LED is installed beneath each playing surface. The LEDs provide immediate visual feedback and support an interactive learning mode based on MIDI scores. The two colors can be used to distinguish notes assigned to the left and right hands. During guided playback, the LED associated with the next expected note is illuminated, and the sequence advances only after the corresponding playing surface has been struck. This provides a simple self-paced method for learning musical sequences directly on the instrument. Moreover, as in the arduino version, the ESP32 can also be used as a simple MIDI-USB controller to send music events through the MIDI protocol.

\begin{figure}[!ht] 
\centering \includegraphics[width=\textwidth]{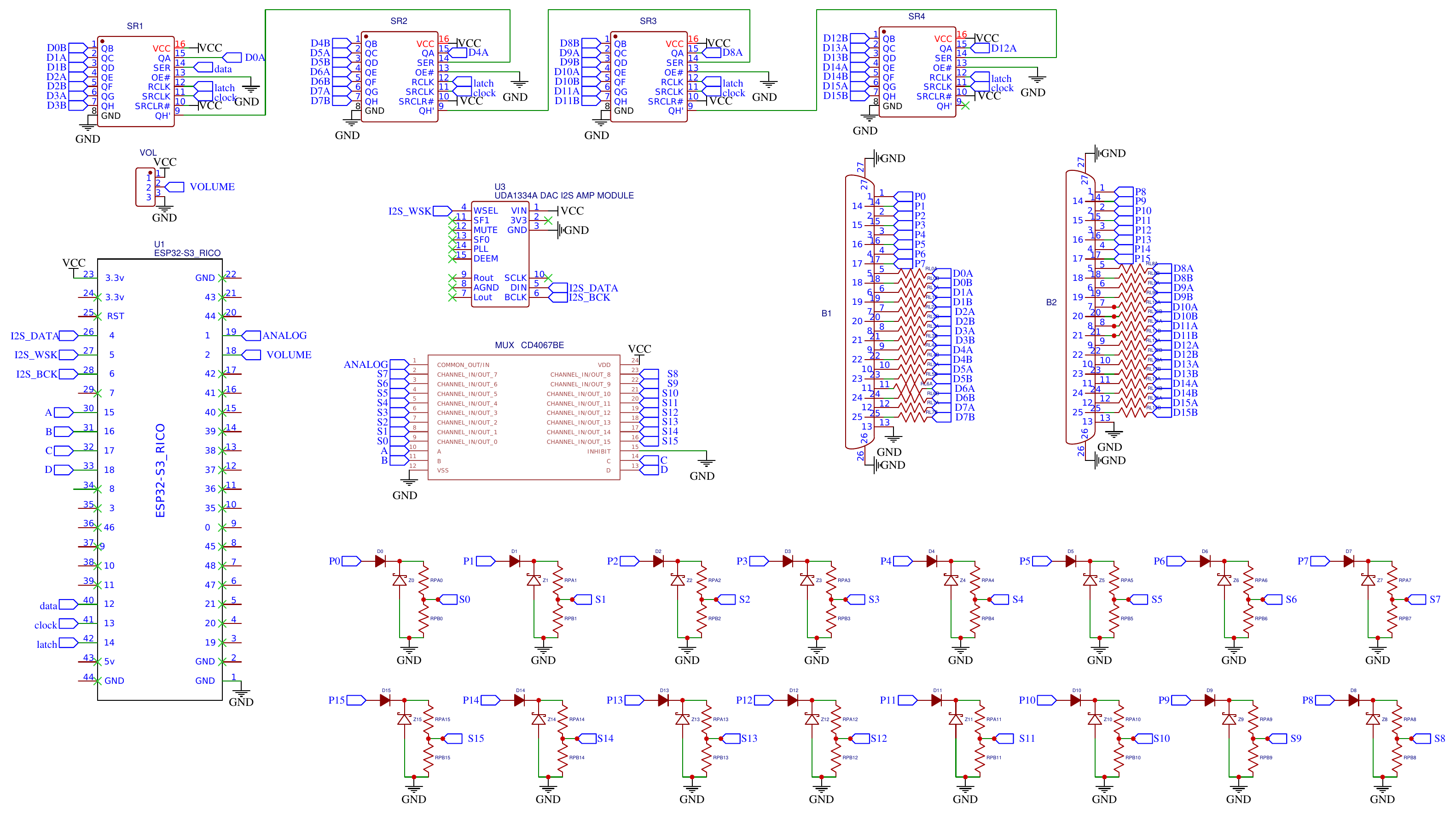}
\caption{Electronic schematic of the ESP32 interface board, including the piezoelectric sensor-conditioning circuits, the CD4067 analog multiplexer, and the cascaded 74HC595 shift registers used to control the bi-color LEDs.} \label{fig:esp32interface} 
\end{figure}

\begin{figure}[!ht] 
\centering 
\subfigure[front]{\includegraphics[width=0.48\textwidth]{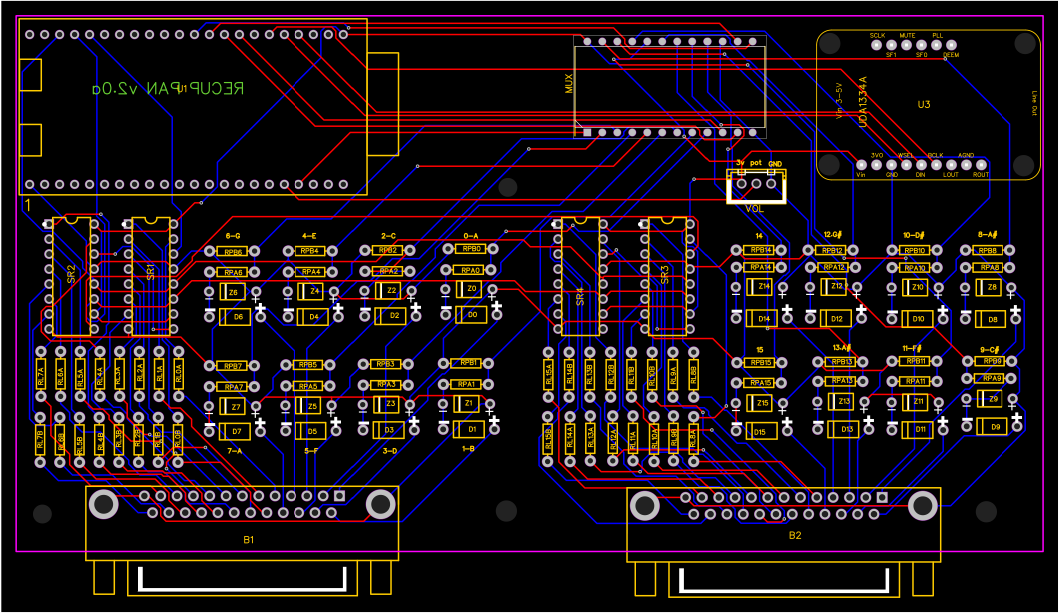}}~
%\subfigure[bottom]{\includegraphics[width=0.45\textwidth]{figs/pcb2.pdf}}\\
\subfigure[3D view]{\includegraphics[width=0.51\textwidth]{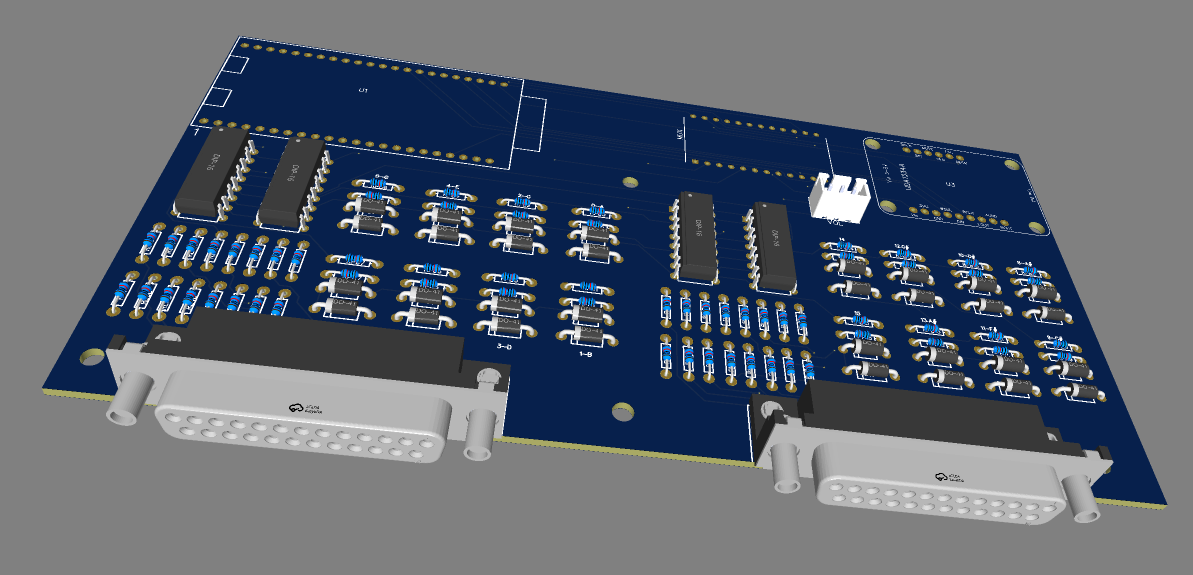}}
\caption{Illustrations of the implemented PCB interface used by the ESP32 microcontroller.} \label{fig:esp32interface2}
\end{figure}

Figs.~\ref{fig:esp32interface} and~\ref{fig:esp32interface2} present the detailed schematic and the implemented \ac{pcb} of the ESP32 interface board. The board connects the salad bowls through a DB25 connector and integrates the piezoelectric signal-conditioning circuits, the CD4067 analog multiplexer, the cascaded 74HC595 shift registers, the UDA1334A audio interface, and the LED current-limiting resistors. Compared with the Arduino implementation, this dedicated PCB increases the hardware integration while considerably simplifying the wiring and assembly of the complete instrument. The complete Gerber files, and bill of materials are provided in the project repository to facilitate reproducibility and further hardware developments.

%-------------------------
%-------------------------

%% file: software.tex
Both hardware implementations rely on dedicated open-source firmware developed
specifically for the \ehandpan{}. Although the Arduino and ESP32 versions
share the same playing interface and sensing principle, their firmware differs
to account for their respective hardware capabilities and intended use. In both
cases, the software continuously acquires the conditioned piezoelectric
signals, detects note impacts, estimates the strike velocity, and associates
each detected event with the corresponding musical note. The firmware also
controls the bi-color LEDs located beneath each playing surface to provide
visual feedback and implement interactive learning modes.

\subsection{Arduino version}

The Arduino firmware is available in the public GitHub repository
\url{https://github.com/dfourer/eHandPan/arduino}. It implements the real-time acquisition of
the piezoelectric sensors and converts detected impacts into
velocity-sensitive MIDI messages. The same source code supports both the
Arduino Uno and Arduino Mega through conditional compilation.

On the Arduino Uno, six piezoelectric sensors are connected directly to analog
inputs A0--A5. On the Arduino Mega, 14 sensors are acquired through
analog inputs A0--A13. Therefore, no analog multiplexer is required in either
configuration. Each analog channel is sampled sequentially within the main
program loop, which is repeated approximately every millisecond.

Impact detection is based on a fixed amplitude threshold. When the value
measured on a channel exceeds the detection threshold and the corresponding
note is not already active, the firmware generates a MIDI Note On message. The
MIDI velocity is obtained by linearly mapping the measured 10-bit analog value
to the MIDI velocity range from 0 to 127. Although this simple method does not
search for the complete peak of the piezoelectric waveform, it provides a
computationally inexpensive estimate of the strike intensity.

To prevent a single mechanical impact from producing several successive MIDI
events, an independent hold-off counter is maintained for each sensor. After a
Note On event, the corresponding note remains active for a minimum duration of
80~ms. When this duration expires, the firmware transmits a MIDI Note Off
message with a velocity of zero. This mechanism acts as a simple retrigger
protection and limits multiple detections caused by the oscillatory response of
the piezoelectric sensors.

The MIDI pitch assigned to each playing surface is defined in a lookup table.
Separate note mappings are provided for the six-note Arduino Uno configuration
and the fourteen-note Arduino Mega configuration. A global transposition
parameter can additionally shift all transmitted notes by a fixed number of
semitones.

MIDI messages are transmitted directly through the hardware serial port as
three-byte MIDI commands. During normal operation, the serial interface is
configured at 31,250~bit/s for compatibility with the MocoLUFA USB-MIDI
firmware. MocoLUFA allows the Arduino board to be recognized by the host
computer as a USB-MIDI device, enabling its use with software synthesizers,
digital audio workstations, and other MIDI-compatible applications.

Two optional diagnostic modes are included in the source code. The first
outputs the estimated velocity of each analog channel as comma-separated data
at 115,200~bit/s, allowing the sensor responses to be monitored and recorded.
The second measures the execution time of each acquisition loop using the
microcontroller timer. The built-in LED connected to digital pin~13 is also
activated when an impact is detected, providing a simple visual indication
during development and testing.

The current Arduino firmware is limited to sensor acquisition and outgoing
MIDI generation. Control of the red/green LEDs installed beneath the playing
surfaces and interpretation of incoming educational MIDI sequences require an
additional firmware module or a separate version of the program.

% Tutorial MocoLufa : https://www.youtube.com/watch?v=-bCz2I9SMAA
% 

\subsection{ESP32 version}

The ESP32 firmware is also available from the public GitHub repository
\url{https://github.com/dfourer/eHandPan/esp32}. Unlike the Arduino implementation, it implements all the functions required by a standalone electronic musical instrument, including sensor acquisition, sample-based audio synthesis, LED control, wireless communication, browser-based configuration, and persistent storage of user settings.

The firmware developed using the PlatformIO framework is organized as a set of cooperative software modules running on top of the ESP32 Arduino framework. A dedicated task continuously scans the piezoelectric sensors independently of the main application loop, ensuring deterministic acquisition despite the concurrent execution of audio synthesis, web services, and user-interface functions. Detected impacts are transferred to the main application through an inter-task queue, where velocity estimation, note generation, and audio playback are performed. This architecture minimizes acquisition latency while preserving responsive operation of the embedded web server and audio engine.

Because the ESP32 exposes an insufficient number of analog inputs for the 14 piezoelectric sensors, the firmware sequentially scans a 16-channel CD4067 analog multiplexer. Each detected impact is validated, its peak amplitude is estimated from successive ADC measurements, and the corresponding MIDI velocity is computed before triggering polyphonic playback of prerecorded handpan samples stored in the ESP32 Flash memory. Audio samples are mixed in real time and streamed through the I$^2$S peripheral to an external UDA1334A digital-to-analog converter, providing low-latency audio reproduction.

The CD4067 analog multiplexer is controlled using the open-source \texttt{CD74HC4067} library, which provides the channel-selection functions used by the firmware to sequentially acquire the fourteen piezoelectric sensors through a single ADC input.

The firmware also controls the four cascaded 74HC595 shift registers driving the fourteen bi-color LEDs. The open-source \texttt{ShiftRegister74HC595} library is used to simplify communication with the cascaded shift registers, allowing the 28 LED channels to be controlled using only three GPIO signals
(DATA, CLOCK, and LATCH). These LEDs provide immediate visual feedback during normal playing and support the embedded educational mode by displaying the next notes extracted from MIDI files.

Instrument parameters, including piezoelectric thresholds and playback settings, are stored in the internal EEPROM to preserve user preferences across power cycles. In addition to its standalone operation, the firmware can also be compiled as a class-compliant USB-MIDI device using TinyUSB,
allowing the instrument to communicate with external digital audio workstations and software synthesizers.

% The firmware also controls the four cascaded 74HC595 shift registers driving the fourteen bi-color LEDs. The open-source \texttt{ShiftRegister74HC595} library is used to simplify communication with the cascaded shift registers while minimizing the number of required GPIOs. These LEDs provide immediate visual feedback during normal playing and support the embedded educational mode by displaying the next notes extracted from MIDI files. Instrument parameters, including piezoelectric thresholds and playback settings, are stored in the internal EEPROM to preserve user preferences across power cycles. In addition to its standalone operation, the firmware can also be compiled as a class-compliant USB-MIDI device using TinyUSB, allowing the instrument to communicate with external digital audio workstations and software synthesizers.

\begin{figure}[!ht]
    \centering
    \includegraphics[width=0.55\textwidth]{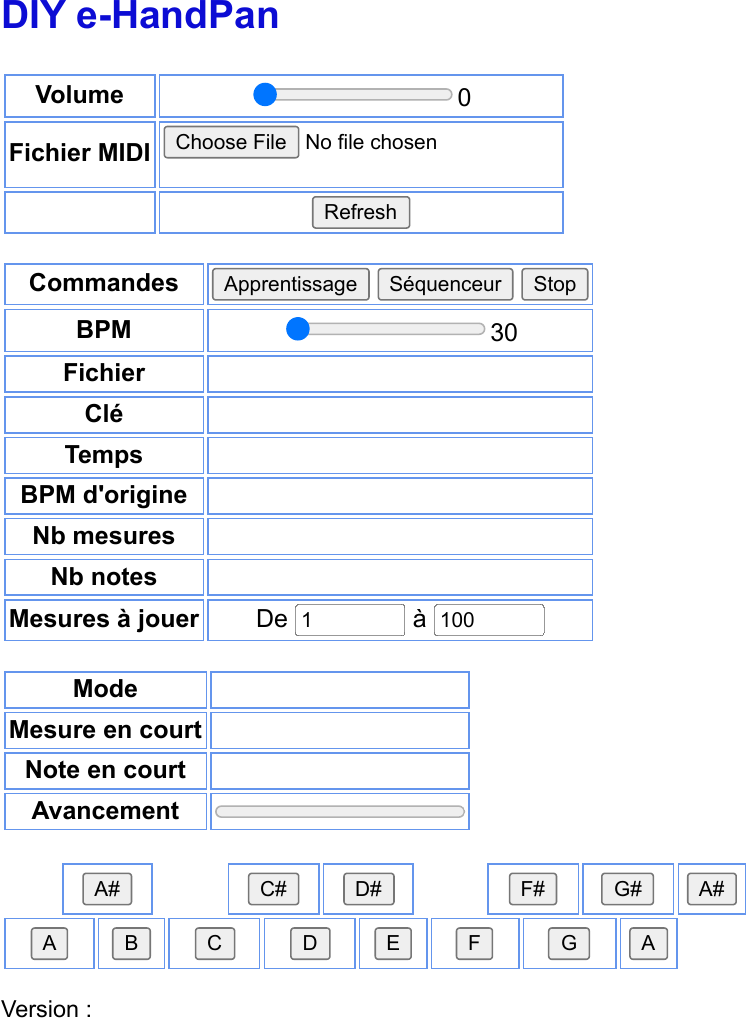}
    \caption{Web-based user interface of the ESP32 implementation. The interface
    is accessible from a standard web browser through the Wi-Fi access point
    created by the instrument and provides access to configuration, playback,
    and educational functions.}
    \label{fig:web_interface}
\end{figure}

Finally, the ESP32 simultaneously hosts a Wi-Fi access point and an asynchronous HTTP server. Through a standard web browser, users can configure the instrument, upload MIDI files, adjust acquisition and playback parameters, monitor diagnostic information, and update the firmware over the air using the integrated ElegantOTA service, without installing dedicated software. The browser-based user interface, illustrated in Fig.~\ref{fig:web_interface}, provides centralized access to these configuration and monitoring functions from a computer, tablet, or smartphone connected to the Wi-Fi network created by the instrument.

%% file: implementation.tex
The construction of the salad bowl, compact-disc striking surfaces, piezoelectric sensors and bi-color LEDs is common to both versions of the instrument. The main difference between the two implementations concerns the electronic interface: the Arduino version uses manually wired components, whereas the ESP32 version requires the dedicated printed circuit board and the DB25 interconnection.

In addition to the step-by-step instructions provided in this section,
video tutorials illustrating the main assembly and construction procedures
are available through the French project YouTube channel\footnote{\url{https://www.youtube.com/playlist?list=PL_k3PczvzYDb-f5t7U6dC4wk6cq4hSw6F}}.
These videos provide complementary visual guidance for the mechanical assembly,
sensor installation, wiring, soldering, and final integration of the instrument.

\subsection{Preparation of the salad bowl}

The first construction step consists in determining the position of each playable note on the outer surface of the salad bowl. Separate layouts are provided for the six-note and eight-note configurations. Each note position is located approximately 2 cm from the center of the corresponding compact disc to accommodate the hole required for the wires, and should therefore be marked before drilling.
The note positions should be distributed evenly around the bowl, leaving sufficient space between adjacent compact discs and avoiding areas with excessive surface curvature.

One hole is drilled for each note. This hole is used to route the electrical conductors associated with both the piezoelectric sensor and the corresponding bi-color LED from the outside of the bowl to the internal electronic connections. The hole diameter must therefore be large enough for the required conductors, while remaining smaller than the compact disc so that it is completely hidden after assembly. Before drilling, the marked positions should be checked against the note-placement template and against the internal clearance of the bowl.

Appropriate eye protection must be worn during drilling. The bowl must be firmly immobilized to prevent rotation or slipping. Any sharp edges produced by drilling should be removed using a deburring tool or fine abrasive paper to avoid damaging the insulation of the electrical wires.

Each compact disc is used as a rigid striking surface. A piece of compliant foam is cut to approximately match the area of the compact disc. A central opening or recess is prepared in the foam to accommodate the piezoelectric sensor. The piezoelectric element is positioned at the center of the foam layer so that mechanical impacts applied to the compact disc are efficiently transmitted to the sensor while reducing direct stress on the ceramic element.

The recommended mechanical stack, from the outer surface to the salad bowl, is:
\begin{enumerate}
    \item compact disc used as the striking surface;
    \item piezoelectric sensor positioned at the center;
    \item shaped foam layer providing mechanical support and partial vibration isolation;
    \item salad-bowl surface.
\end{enumerate}

The bi-color LED is installed close to the corresponding compact disc so that the illuminated note can be identified by the player. Its conductors are routed through the same drilled hole as the piezoelectric sensor wires. %% DF : si il n y a qu un seul trou par note qui sert au piezo + leds

The complete assembly is attached to the salad bowl using a strong multipurpose adhesive or high-strength double-sided adhesive tape. The selected adhesive must provide sufficient bonding strength while remaining compatible with the plastic or metallic surface of the bowl, the foam material and the compact disc. When liquid adhesive is used, the assembly must remain immobilized until the adhesive is fully cured. Excess adhesive must not be allowed to reach the piezoelectric ceramic element or the electrical connections.

The piezoelectric conductors should be kept as short as reasonably possible and should not be subjected to mechanical tension. A small amount of strain relief may be added on the internal side of the bowl using adhesive tape or a drop of flexible adhesive. All note assemblies should be tested electrically before the final electronic enclosure is closed.

%%%%%%%%%%%%%%%%%%%%%%%%%%%%%%%%%%%%%%%%%%%%%%%%%%%%%%%%%%%%%%%%%%%%%%%%%%%%%%%%%%%%%%%%%%%%%%

\subsection{Arduino version}

The Arduino implementation is intended as the simplest version to reproduce without manufacturing a dedicated printed circuit board. The six-note configuration uses an Arduino Uno, whereas the extended configuration uses an Arduino Mega to provide the additional analog inputs and digital outputs required by the instrument.

Each piezoelectric sensor is connected to one analog input through the dedicated input-protection and signal-conditioning circuit shown in the corresponding schematic. Each bi-color LED is connected to two digital output pins through current-limiting resistors. A resistance of \(100~\Omega\) is connected in series with each LED color channel.

For this version, the electronic components are soldered manually onto the connecting wires. The resistors, protection components and interconnections are assembled using point-to-point wiring, solder joints and heat-shrink tubing. All exposed electrical joints must be insulated to prevent short circuits between adjacent conductors or against the bowl when a conductive metallic bowl is used.

The connection procedure for the Arduino version is as follows:
\begin{enumerate}
    \item identify and label the two conductors of each piezoelectric sensor;
    \item identify the common terminal and the two color terminals of each bi-color LED;
    \item connect each piezoelectric sensor to its corresponding input-protection circuit;
    \item connect each conditioned piezoelectric signal to the assigned Arduino analog input;
    \item connect both LED color channels to their assigned Arduino digital outputs through \(100~\Omega\) series resistors;
    \item connect all required ground conductors to the Arduino ground;
    \item verify electrical continuity and the absence of short circuits before powering the board.
\end{enumerate}

The correspondence between the note names, MIDI note numbers, analog inputs and LED output pins is defined in the supplied Arduino firmware. The same source code is used for both the Arduino Uno and Arduino Mega versions. The target board selected in the Arduino IDE or in \texttt{arduino-cli} determines the hardware configuration compiled for the corresponding microcontroller.

For the six-note Arduino Uno version, the implemented notes are:

\begin{verbatim}
Bb3, Db4, Eb4, Gb4, Ab4, Bb4
\end{verbatim}

For the fourteen-note Arduino Mega version, the implemented notes are:

\begin{verbatim}
Bb3, Db4, Eb4, Gb4, Ab4, Bb4,
A3, B3, C4, D4, E4, F4, G4, A4
\end{verbatim}

\subsubsection{Uploading and testing the Arduino firmware}

Before replacing the original USB-to-serial firmware with MocoLUFA, the eHandPan sketch must be uploaded to the Arduino main microcontroller. The source code is provided in the project GitHub repository and can be compiled and uploaded using the Arduino IDE or \texttt{arduino-cli}.

For the Arduino Uno, the target board is selected as:

\begin{verbatim}
Arduino Uno
\end{verbatim}

For the Arduino Mega, the target board is selected as:

\begin{verbatim}
Arduino Mega or Mega 2560
\end{verbatim}

The sketch should be uploaded while the original Arduino USB-to-serial firmware is still installed on the ATmega16U2. At this stage, the serial connection remains accessible through the conventional virtual serial port and can therefore be used to verify the hardware and firmware operation.

Each piezoelectric sensor should first be tested individually by striking the associated compact-disc surface and checking that the expected MIDI message is transmitted. The response of each bi-color LED should also be verified by sending the corresponding MIDI notes back to the Arduino.

Because the Arduino sketch transmits raw MIDI bytes through its hardware serial interface, the communication can be tested using a serial-to-MIDI bridge such as Hairless MIDI Serial Bridge. Hairless receives the MIDI byte stream from the Arduino serial port and forwards it to a virtual MIDI port that can be connected to a software synthesizer, sampler or digital audio workstation.

The serial-to-MIDI test configuration is:
\begin{enumerate}
    \item upload the supplied eHandPan sketch to the Arduino;
    \item close the Arduino serial monitor, since only one application can normally access the serial port at a time;
    \item start Hairless MIDI Serial Bridge;
    \item select the Arduino serial port;
    \item configure the serial communication rate expected by the firmware;
    \item connect the Hairless MIDI output to a virtual MIDI port or directly to the selected MIDI application;
    \item strike each note and verify the received pitch and velocity;
    \item send MIDI notes toward the Arduino and verify the corresponding LED behavior.
\end{enumerate}

This intermediate test stage is strongly recommended because it allows wiring errors, incorrect pin assignments, defective piezoelectric sensors and LED polarity errors to be identified before modifying the USB-interface firmware.

The serial bridge may also be used temporarily to operate the instrument. However, this configuration introduces additional communication and software-routing latency. For satisfactory real-time musical performance, the ATmega16U2 must subsequently be programmed with MocoLUFA, which exposes the Arduino directly as a class-compliant USB-MIDI interface and removes the need for the intermediate serial-to-MIDI bridge.

A future version of the Arduino interface is planned around a dedicated shield. This shield will integrate the input-protection components, LED resistors and signal-routing connections on a printed circuit board. It will also provide a DB25 connector to simplify the connection between the salad bowl and the Arduino electronics. This future implementation is intended to replace point-to-point soldering while remaining electrically compatible with the current Arduino firmware.

\subsubsection{Installing the USB-MIDI firmware}

After the Arduino sketch and all instrument functions have been verified through the conventional serial interface, the USB interface microcontroller can be reprogrammed with open-source MocoLUFA\footnote{\url{https://github.com/kuwatay/mocolufa}}.

MocoLUFA transforms the USB interface of Arduino Uno and Mega boards into a class-compliant USB-MIDI device. This direct USB-MIDI implementation eliminates the need for external serial-to-MIDI bridge software and provides a low-latency, reliable communication channel suitable for real-time musical performance.

Unlike the eHandPan sketch, which is stored in the main microcontroller---the ATmega328P on the Arduino Uno or the ATmega2560 on the Arduino Mega---MocoLUFA is installed in the dedicated USB interface microcontroller. Recent official Arduino Uno and Mega boards use an ATmega16U2 for this purpose, whereas some older revisions use an ATmega8U2.

\begin{figure}[!ht]
    \centering
    \includegraphics[width=0.4\textwidth]{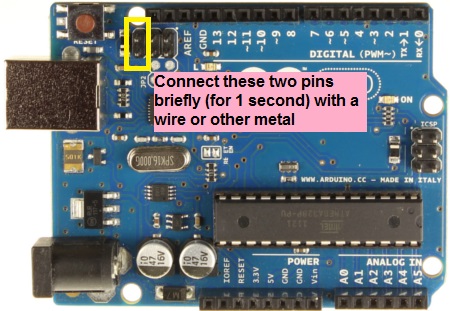}
    \caption{RESET and GND pins of the ATmega16U2 used to place the USB interface microcontroller into DFU mode before installing the MocoLUFA firmware.}
    \label{fig:mocolufa}
\end{figure}

Installing MocoLUFA does not erase or modify the eHandPan sketch stored in the main microcontroller. Only the firmware responsible for USB communication with the host computer is replaced. Before programming, the Arduino must be placed into Device Firmware Upgrade (DFU) mode by briefly shorting the RESET and GND pins of the USB interface microcontroller, as illustrated in Fig.~\ref{fig:mocolufa}.

%-------------------------------------------------------------------------
\begin{figure}[!ht]
    \centering
    \includegraphics[width=0.4\textwidth]{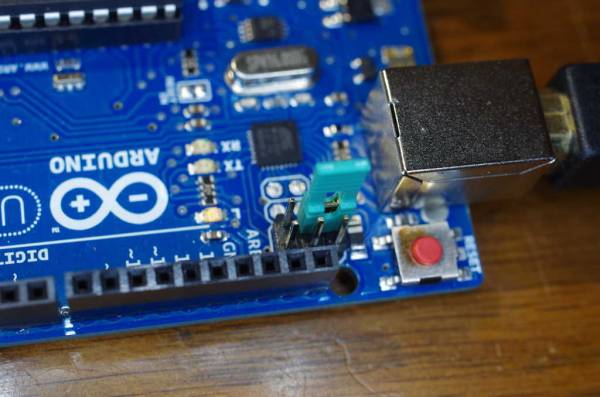}
    \caption{Pins to be connected with a jumper to switch the Arduino from USB-MIDI mode back to the standard USB serial interface. This operation allows a new sketch to be uploaded without reinstalling the original USB interface firmware.}
    \label{fig:mocolufa-serial}
\end{figure}
%-------------------------------------------------------------------------

Under Linux, the required DFU programming utility can be installed using:

\begin{terminal}
sudo apt install dfu-programmer
\end{terminal}

The MocoLUFA source code is downloaded and compiled using:

\begin{terminal}
git clone https://github.com/kuwatay/mocolufa.git
cd mocolufa/Arduino-usbserial
make clean
make
\end{terminal}

The compilation generates the \texttt{Arduino-usbserial.hex} firmware file.

To program this file, the ATmega16U2 must first be placed in Device Firmware Upgrade (DFU) mode. With the Arduino connected and powered through USB, briefly short the RESET and GND pins associated with the ATmega16U2, located near the USB connector. Only these two pins must be shorted, and the connection must be removed immediately after the reset.

The presence of the device in DFU mode can be verified using:

\begin{terminal}
sudo dfu-programmer atmega16u2 get
\end{terminal}

MocoLUFA is then installed using:

\begin{terminal}
sudo dfu-programmer atmega16u2 erase
sudo dfu-programmer atmega16u2 flash Arduino-usbserial.hex
sudo dfu-programmer atmega16u2 reset
\end{terminal}

For older Arduino boards equipped with an ATmega8U2, the device identifier \texttt{atmega16u2} must be replaced with \texttt{atmega8u2} in all \texttt{dfu-programmer} commands.

Once installed, MocoLUFA converts the UART byte stream generated by the DIY eHandPan sketch into standard USB-MIDI messages. Consequently, the serial communication between the main Arduino microcontroller and the USB interface must use the standard MIDI baud rate of \(31\,250\)~baud. In the supplied firmware, this is implemented as:

\begin{verbatim}
Serial.begin(31250);
\end{verbatim}

This value must not be modified. Likewise, no diagnostic messages, debugging information or other ASCII text should be transmitted over the same serial interface while MocoLUFA is active, since every transmitted byte is interpreted as MIDI data.

After the DFU programming procedure, the USB cable should be disconnected and reconnected.
The Arduino should then be recognized by the operating system as a class-compliant USB-MIDI device instead of a virtual serial port. The device can be selected directly in any digital audio workstation or MIDI-compatible software, making Hairless MIDI Serial Bridge unnecessary. Incoming MIDI messages can trigger the virtual instrument, while outgoing messages generated by the software can also be routed back to the DIY eHandPan to drive the bi-color LEDs.

The device can be selected directly as a MIDI input and output in the digital audio workstation. Incoming MIDI messages are used to trigger the virtual instrument, while outgoing MIDI messages can be routed back to the eHandPan to control the bi-color LEDs.

If the Arduino sketch must subsequently be modified, the board can be temporarily switched back to its conventional USB serial mode by connecting the pins with a jumper as shown in Fig.~\ref{fig:mocolufa-serial}. 
The updated sketch can then be uploaded using the Arduino IDE. Once the upload is complete, the board can be switched back to USB-MIDI mode by removing the jumper, without reinstalling the MocoLUFA firmware.

%%%%%%%%%%%%%%%%%%%%%%%%%%%%%%%%%%%%%%%%%%%%%%%%%%%%%%%%%%%%%%%%%%%%%%%%%%%%%%%%%%%%%%%%%%%%%%%%%%%%

\subsection{ESP32 version}

\subsubsection{Building the PCB}

% \begin{figure}[!ht]
%  \centering
%  \includegraphics[width=0.8\textwidth]{figs/DB25.pdf}
% \caption{DB25 connector pin assignment for the piezoelectric sensors and bi-color LEDs used by the ESP32 interface. As a mnemonic, the right-hand notes are assigned to the red color (R for Right and Red).}
%  \label{fig:esp32-db25}
% \end{figure}

\begin{table}[!ht]
\centering
\caption{DB25 pin assignment between the salad bowl and the ESP32 interface board. Pins 26 and 27 are connected to ground. Both the 6-note and 8-note versions use the same DB25 connector. Signals denoted $Pxx$ correspond to piezoelectric sensor inputs, whereas signals denoted $Dxx$ correspond to the bi-color LED connections.}
\label{tab:db25}
\small
%\resizebox{0.45\textwidth}{!}{
\begin{tabular}{clll}
\hline
Pin & ESP32 signal & 8-note instrument & 6-note instrument \\
\hline
1  & P0   & A3      & Bb3 \\
2  & P2   & C4      & Eb4 \\
3  & P4   & E4      & Ab4 \\
4  & P6   & G4      & -- \\
5  & D0A  & A3 Left & Bb3 Left \\
6  & D1A  & B3 Left & Db4 Left \\
7  & D2A  & C4 Left & Eb4 Left \\
8  & D3A  & D4 Left & Gb4 Left \\
9  & D4A  & E4 Left & Ab4 Left \\
10 & D5A  & F4 Left & Bb4 Left \\
11 & D6A  & G4 Left & -- \\
12 & D7A  & A4 Left & -- \\
13 & GND  & Ground & Ground \\
14 & P1   & B3      & Db4 \\
15 & P3   & D4      & Gb4 \\
16 & P5   & F4      & Bb4 \\
17 & P7   & A3      & -- \\
18 & D0B  & A3 Right & Bb3 Right \\
19 & D1B  & B3 Right & Db4 Right \\
20 & D2B  & C4 Right & Eb4 Right \\
21 & D3B  & D4 Right & Gb4 Right \\
22 & D4B  & E4 Right & Ab4 Right \\
23 & D5B  & F4 Right & Bb4 Right \\
24 & D6B  & G4 Right & -- \\
25 & D7B  & A4 Right & -- \\
26 & GND  & Ground & Ground \\
27 & GND  & Ground & Ground \\
\hline
\end{tabular} %}
%~ \includegraphics[width=0.45\textwidth]{figs/pcbdb25.png}
\end{table}
% 
% \begin{figure}
%     \centering
%     \includegraphics[width=0.6\textwidth]{figs/pcbdb25.png}
%     \caption{DB25 pin assignment between the salad bowl and the ESP32 interface board. Pins 26 and 27 are connected to ground. Both the 6-note and 8-note versions use the same DB25 connector. The ESP32 signal can correspond to piezo denoted $Pxx$ or LED denoted $Dxx$.}
%     \label{fig:pcbdb25}
% \end{figure}

The ESP32 implementation requires the dedicated \acf{pcb} (Figs.~\ref{fig:esp32interface} and \ref{fig:esp32interface2}). Unlike the Arduino implementation, point-to-point wiring is not recommended because the interface combines multiplexed analog acquisition, serial LED control, audio generation and multiple power and ground connections on a compact printed circuit board.

The salad bowl is connected to the \ac{pcb} through the DB25 connector. The wiring must strictly follow Table~\ref{tab:db25}. Before soldering the connector, each conductor should be labelled according to its note position and function (sensor, LED, and ground connections).

The recommended assembly procedure for the ESP32 version is as follows:
\begin{enumerate}
    \item prepare and label all conductors originating from the salad bowl;
    \item check the correspondence between notes and conductors;
    \item route the piezoelectric sensor and LED conductors to the DB25 connector;
    \item solder each conductor to the corresponding DB25 pin listed in Table~\ref{tab:db25};
    \item inspect all solder joints;
    \item check the continuity between the salad bowl and the PCB terminals;
    \item check the the absence of short circuits between adjacent DB25 pins;
    \item connect the DB25 cable to the PCB only after all electrical tests have passed.
\end{enumerate}

The PCB integrates the piezoelectric signal-conditioning circuits, the CD4067 analog multiplexer, the 4 cascaded 74HC595 shift registers, the ESP32-S3 module and the audio-output circuitry. All components must be assembled according to the PCB fabrication files and bill of materials provided in the project repository. Particular attention should be paid to polarized components, integrated-circuit orientation and connector placement.

The ESP32 board must not be powered while inserting or removing the DB25 connector. Incorrect DB25 wiring may expose ESP32 inputs or LEDs to unsuitable voltages and may permanently damage the PCB. For this reason, the completed cable must be checked with a multimeter before its first connection to the powered instrument.

After PCB assembly, the ESP32 firmware is uploaded through the standard USB programming interface. The firmware controls analog acquisition through the CD4067 multiplexer, drives the LEDs using the 74HC595 shift register, generates or streams audio through the I$^2$ interface and provides the network-based control functions described in the software section.

\subsubsection{Flashing the firmware}

The firmware is developed using the PlatformIO framework, which is supported on Linux, Windows, and macOS. The examples below use the PlatformIO command-line interface under Linux, although equivalent operations can also be performed using the PlatformIO extension for Visual Studio Code to compile and upload the firmware through the standard USB programming interface.

The ESP32-S3 development board includes a factory-installed USB bootloader, so no external programmer is required. All required software libraries are automatically downloaded by PlatformIO during the first compilation.

After cloning the project repository,

\begin{terminal}
git clone https://github.com/dfourer/eHandPan.git
cd eHandPan/esp32
\end{terminal}

the firmware is compiled using

\begin{terminal}
pio run -e esp32_s3_N16R8
\end{terminal}

and uploaded to the board through the USB interface using

\begin{terminal}
pio run -e esp32_s3_N16R8 -t upload
\end{terminal}

During development, the serial output can be monitored using

\begin{terminal}
pio device monitor -e esp32_s3_N16R8
\end{terminal}

Several PlatformIO environments are provided to support different firmware configurations. For example, the environment \texttt{esp32\_s3\_N16R8\_MIDI} enables the USB-MIDI functionality by replacing the environment name in the previous commands.

\subsubsection{First start-up}

After programming, the ESP32 automatically restarts and executes the firmware. The status LED indicates that the initialization procedure has completed successfully before normal operation begins. The firmware then creates its own Wi-Fi access point and starts the embedded asynchronous HTTP server.

Using a standard web browser connected to the ESP32 access point, the user can access the embedded configuration interface described in Section~\ref{sec:software}. 
This interface allows MIDI files to be uploaded, piezoelectric acquisition thresholds and playback parameters to be adjusted, diagnostic information to be monitored, and firmware updates to be performed over the air using the integrated ElegantOTA service.

%% file: validation.tex
The proposed DIY eHandPan was validated at both the subsystem level and the complete instrument level. The experimental evaluation focused on the piezoelectric signal-conditioning circuit, the embedded acquisition system, and the overall functionality of the Arduino and ESP32 implementations.

\subsection{Validation of the piezoelectric conditioning circuit}

\begin{figure}[!ht]
    \centering
    \includegraphics[width=0.5\textwidth]{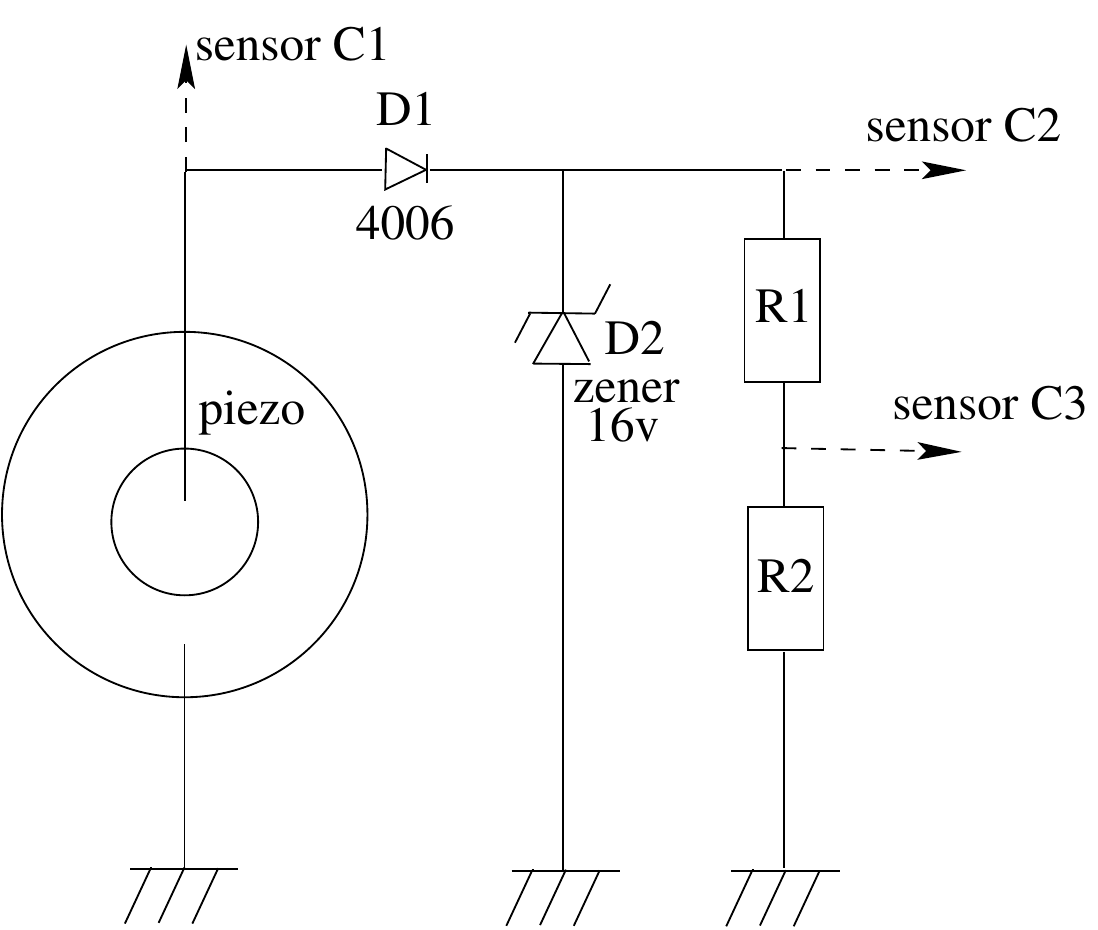}
    \caption{Schema of the evaluation circuit for conditioning the piezoelectric sensor.}
    \label{fig:valid_condition}
\end{figure}

The analog conditioning circuit presented in Section~\ref{sec:conditioning} was designed following a series of experiments aimed at characterizing the electrical response of the piezoelectric sensors under representative playing conditions. To evaluate the effectiveness of the proposed conditioning stage, the experimental setup shown in Fig.~\ref{fig:valid_condition} was assembled using a single piezoelectric sensor.

Oscilloscope measurements were performed at three different locations of the circuit. Channel C1 corresponds to the raw voltage generated by the piezoelectric sensor. Channel C2 measures the signal after the protection stage composed of the Zener diode D1 and the conventional diode D2, which clamp positive overvoltages and negative voltage excursions, respectively. Finally, channel C3 corresponds to the conditioned signal after the voltage divider formed by resistors R1 and R2. This stage attenuates the sensor voltage so that it remains within the input range of the target microcontroller, namely 0--5~V for the Arduino and 0--3.3~V for the ESP32.

\begin{figure}[!ht]
 \centering
     \subfigure[Ali express]{\includegraphics[width=0.48\textwidth]{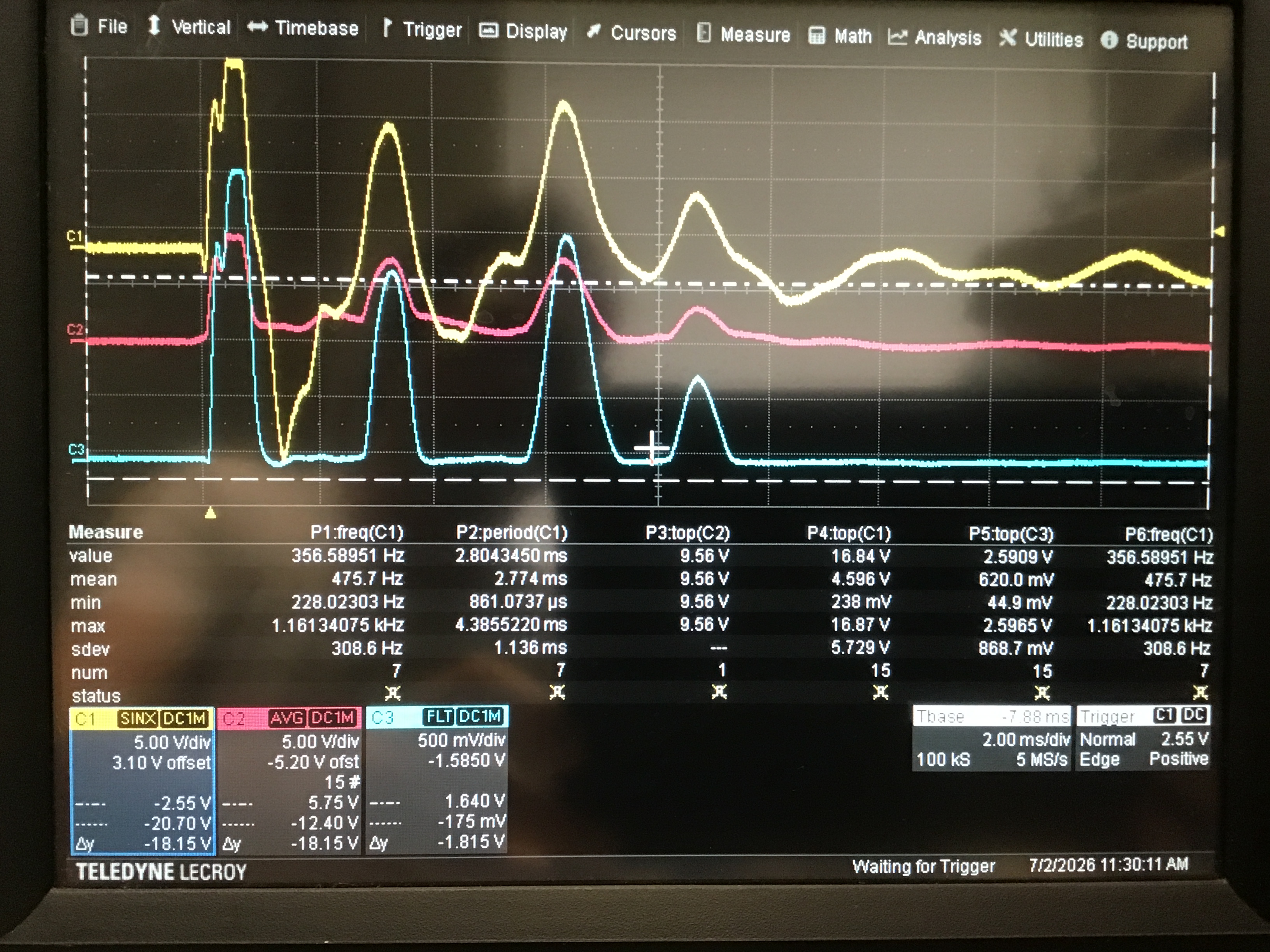}}
     \subfigure[Amazon]{\includegraphics[width=0.48\textwidth]{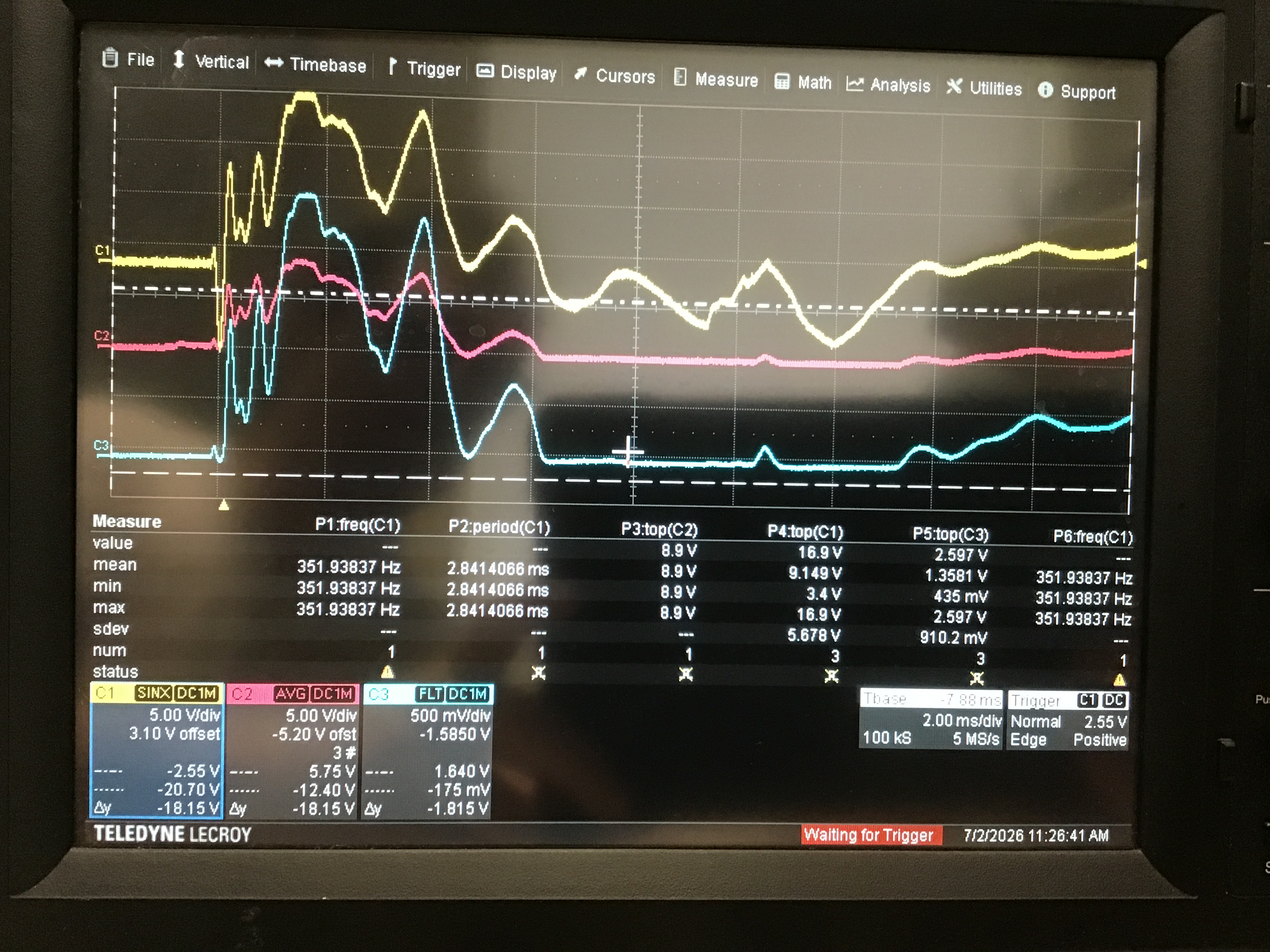}}
     \subfigure[RS component]{\includegraphics[width=0.5\textwidth]{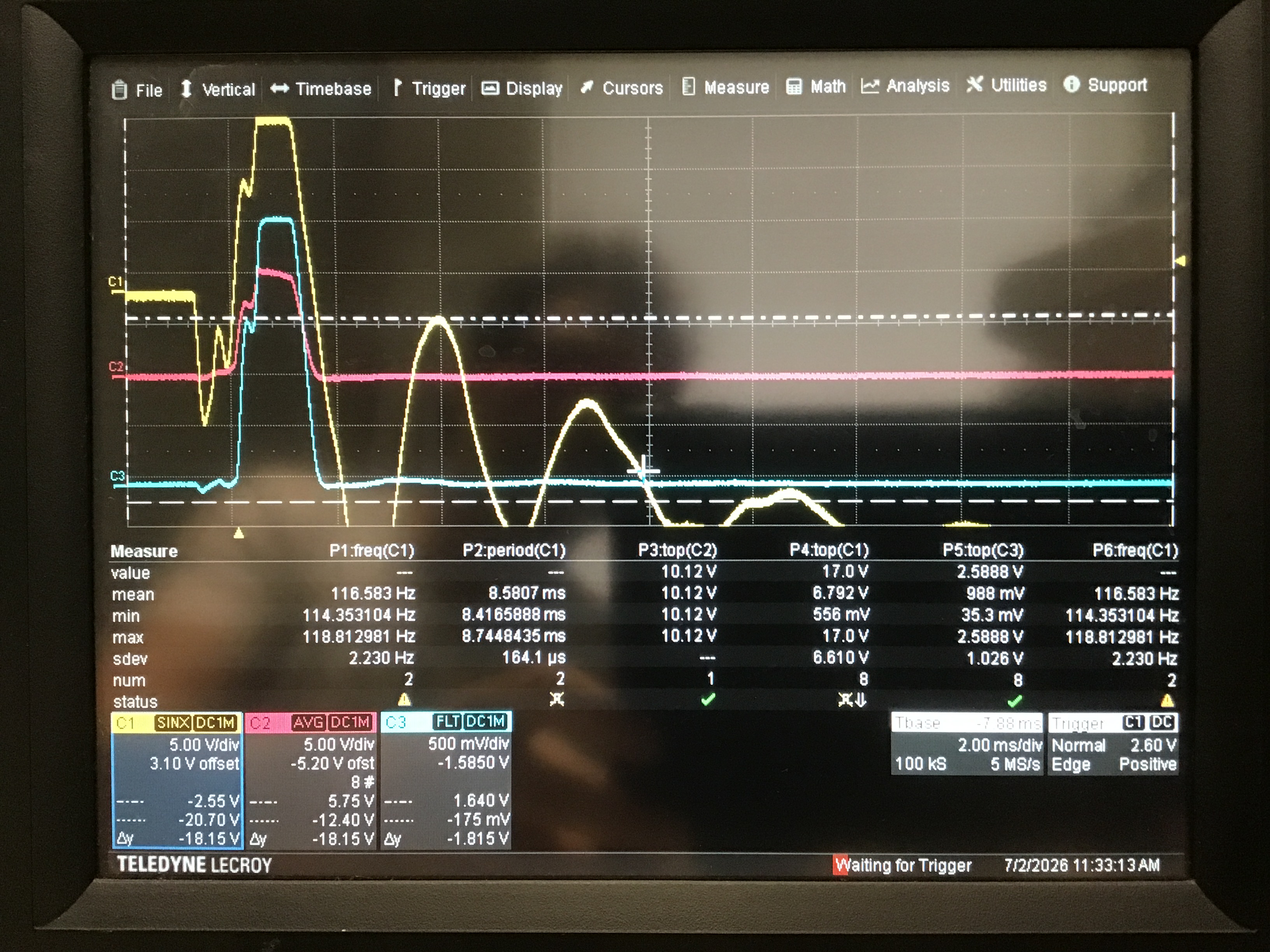}}
     \caption{Oscilloscope measurements obtained with the three commercially available piezoelectric sensors under representative impacts (forte). Channel C1 corresponds to the raw piezoelectric voltage, C2 to the protected signal after the diode stage, and C3 to the conditioned voltage applied to the microcontroller ADC.}
     \label{fig:piezo}
\end{figure}

Representative waveforms obtained with three commercially available piezoelectric sensors are shown in Fig.~\ref{fig:piezo}. In each experiment, the three oscilloscope channels correspond to the measurement points introduced in Fig.~\ref{fig:valid_condition}: C1 is the raw voltage generated by the piezoelectric sensor, C2 is the protected signal after the diode stage, and C3 is the conditioned signal applied to the microcontroller ADC after the voltage divider.

As shown in Fig.~\ref{fig:piezo} and Table.~\ref{tab:piezo_comparison}, the 3 sensors exhibit noticeably different electrical responses despite having similar nominal characteristics. The 27~mm AliExpress sensor (Fig.~\ref{fig:piezo}a) exhibits the lowest effective sensitivity together with the shortest transient response. The 27~mm Amazon sensor (Fig.~\ref{fig:piezo}b) exhibits a more energetic transient response while preserving a waveform similar to that of the previous sensor. Finally, the 50~mm RS Components sensor (Fig.~\ref{fig:piezo}c) exhibits the highest sensitivity together with a longer oscillatory response, resulting in a larger signal duration after the initial impact.

For the three evaluated sensors, the protection stage successfully suppresses the negative voltage excursions and limits the voltage applied to the conditioning network. Furthermore, the voltage divider attenuates the signal to values compatible with the input range of both the Arduino and ESP32 analog-to-digital converters while preserving the transient waveform required for reliable impact detection and MIDI velocity estimation.

These experiments show that commercially available piezoelectric sensors may exhibit significantly different amplitudes and temporal responses. Consequently, the impact detection thresholds implemented in the firmware should be adjusted according to the selected sensor. The proposed conditioning circuit nevertheless provides reliable operation with all evaluated sensors, making the hardware compatible with a wide range of commercially available piezoelectric transducers.

\begin{table}[!ht]
\centering
\caption{Experimental comparison of the evaluated piezoelectric sensors. The peak voltages correspond to the three measurement points defined in Fig.~\ref{fig:valid_condition}: C1 (raw piezoelectric output), C2 (after the protection stage), and C3 (conditioned signal applied to the ADC).}
\label{tab:piezo_comparison}
\resizebox{\textwidth}{!}{\begin{tabular}{|l|c|c|c|c|c|l|}
\hline
\textbf{Piezoelectric sensor} &
\textbf{Diameter} &
\textbf{Peak C1 (V)} &
\textbf{Peak C2 (V)} &
\textbf{Peak C3 (V)} &
\textbf{Sensitivity} &
\textbf{Waveform type} \\
\hline
AliExpress & 27 mm & 16.87 & 9.56 & 2.5965 & Low & Short transient\\
Amazon     & 27 mm & 16.90 & 8.90 & 2.5970 & Medium & Intermediate Response\\
RS Components & 50 mm & 17.00 & 10.12 & 2.5888 & High & Longer oscillatory response \\
\hline
\end{tabular}}
\end{table}

The measurements reported in Table~\ref{tab:piezo_comparison} show that the
protection stage limits the maximum voltage applied to the conditioning circuit
to approximately 9--10~V despite raw piezoelectric voltages approaching
17~V. After attenuation by the voltage divider, the ADC input voltage remains
close to 2.6~V for all evaluated sensors, which is well within the input range
of both the Arduino (5~V) and ESP32 (3.3~V) microcontrollers. These results
experimentally validate the proposed protection and conditioning circuit.

% \begin{table}[ht]
% \centering
% \caption{Experimental characterization of the piezoelectric conditioning circuit.}
% \label{tab:conditioning_results}
% \begin{tabular}{lcc}
% \hline
% Parameter & Measured value & Unit\\
% \hline
% Maximum piezo voltage & XX & V\\
% Maximum ADC input voltage & XX & V\\
% Measured attenuation ratio & XX & --\\
% Theoretical attenuation ratio & XX & --\\
% Maximum clamped voltage & XX & V\\
% \hline
% \end{tabular}
% \end{table}

\subsection{Educational and performance validation}

\begin{figure}[!ht]
    \centering
    \includegraphics[width=0.6\textwidth]{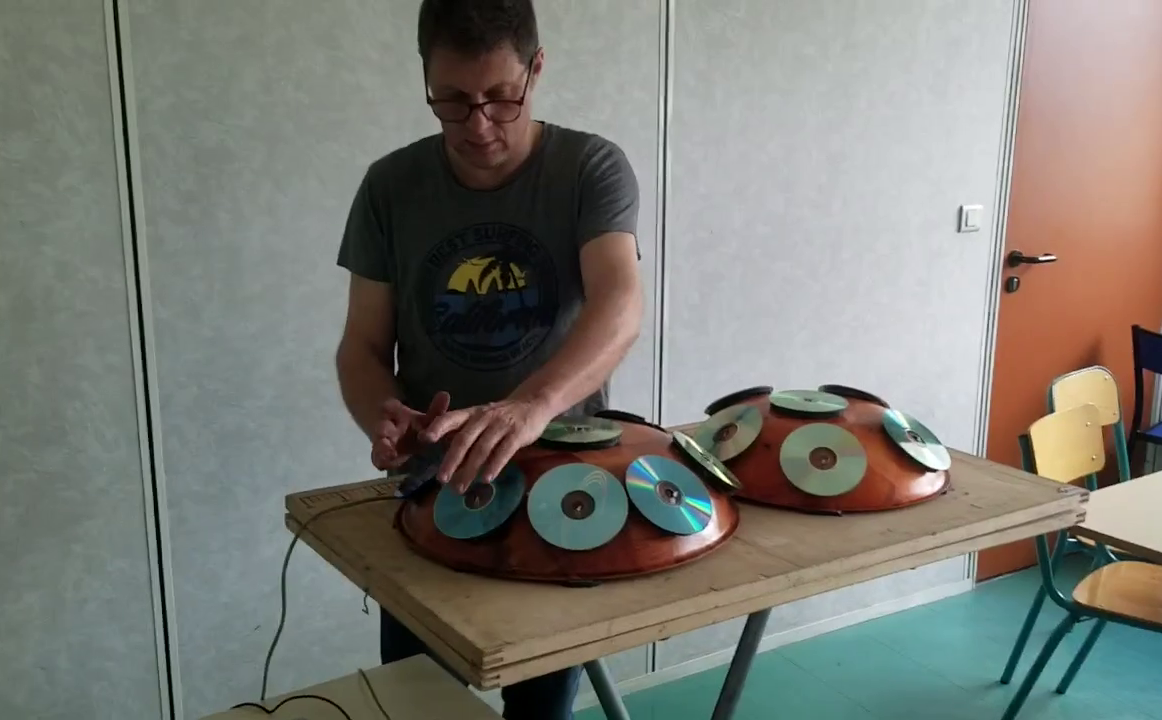}
   \caption{Performance of the fourteen-note Arduino implementation used as a USB-MIDI controller. A demonstration video illustrating the instrument during performance is available at \url{https://youtu.be/22MzmoRZ-XM}.}
    \label{fig:handpandemo}
\end{figure}

Both implementations (Arduino and ESP32) were evaluated over several playing sessions involving repeated impacts on all notes to verify the robustness of the acquisition electronics and the stability of the firmware during continuous operation.
All fourteen playing surfaces were successfully detected during normal playing conditions and the polyphony is supported (cf. Demo presented in Fig.~\ref{fig:handpandemo}).

For the Arduino implementation, the firmware continuously acquired the piezoelectric sensors, estimated the impact velocity, and transmitted USB-MIDI events through the MocoLUFA firmware. The generated MIDI messages were successfully recognized by standard digital audio workstations and software synthesizers.
Informal timing measurements performed on the Arduino Uno implementation showed that the firmware processing loop required less than 10~ms under normal operating conditions. Although this measurement does not include the latency of the external synthesizer or the host computer, it indicates that the embedded acquisition and MIDI generation introduce only a limited delay, which was found to be suitable for real-time musical performance.

For the ESP32 implementation, the firmware continuously scanned the fourteen piezoelectric sensors through the CD4067 analog multiplexer while simultaneously managing sampled audio playback, LED control, USB-MIDI communication, and the embedded web server. The browser-based configuration interface was successfully accessed from desktop and mobile devices connected directly to the Wi-Fi access point generated by the instrument.

The educational mode was also validated using representative MIDI files. During guided playback, the bi-color LEDs correctly indicated the next expected note, and the musical sequence advanced only after the corresponding playing surface had been struck.

Beyond the laboratory evaluation, both implementations were successfully used during educational activities at the University of Evry Paris-Saclay.
A variant of the Arduino version was implemented during a Master's project in 2024 of the Science and Technology
UFR\footnote{\url{https://www.universite-paris-saclay.fr/formation/master/electronique-energie-electrique-automatique/m1-e3a-site-evry}}, in which students assembled, programmed, and evaluated the instrument.
The ESP32 implementation was also subsequently used during practical workshops organized by the Department of Music of the University of Evry Paris-Saclay. These activities confirmed that both versions can be assembled and programmed by students while serving as practical supports for teaching embedded systems, digital audio, and interactive musical
interfaces.